\documentclass[11pt,a4paper]{article}

\usepackage{ascmac}
\usepackage{amsmath}
\usepackage{amssymb}
\usepackage{bm}
\usepackage[footnotesize,bf]{caption}
\usepackage{fullpage}
\usepackage{color}
\usepackage{float}
\usepackage{graphicx}
\usepackage[]{natbib}
\usepackage{subfigure}
\usepackage{txfonts}
\usepackage{threeparttable}
\usepackage{wrapfig}
\usepackage[]{multicol}
\usepackage{wrapfig}
\usepackage{authblk}

\usepackage{floatflt}

\title{\large Complete list of the ASTRO-H Science Working Group}
\date{\vspace{-0.5cm}}
\newcommand{\MakeWhitePaperTitle}{
	\begin{center}
		\begin{figure}
			\vspace{1cm}
			\begin{center}
				\includegraphics[width=0.2\hsize]{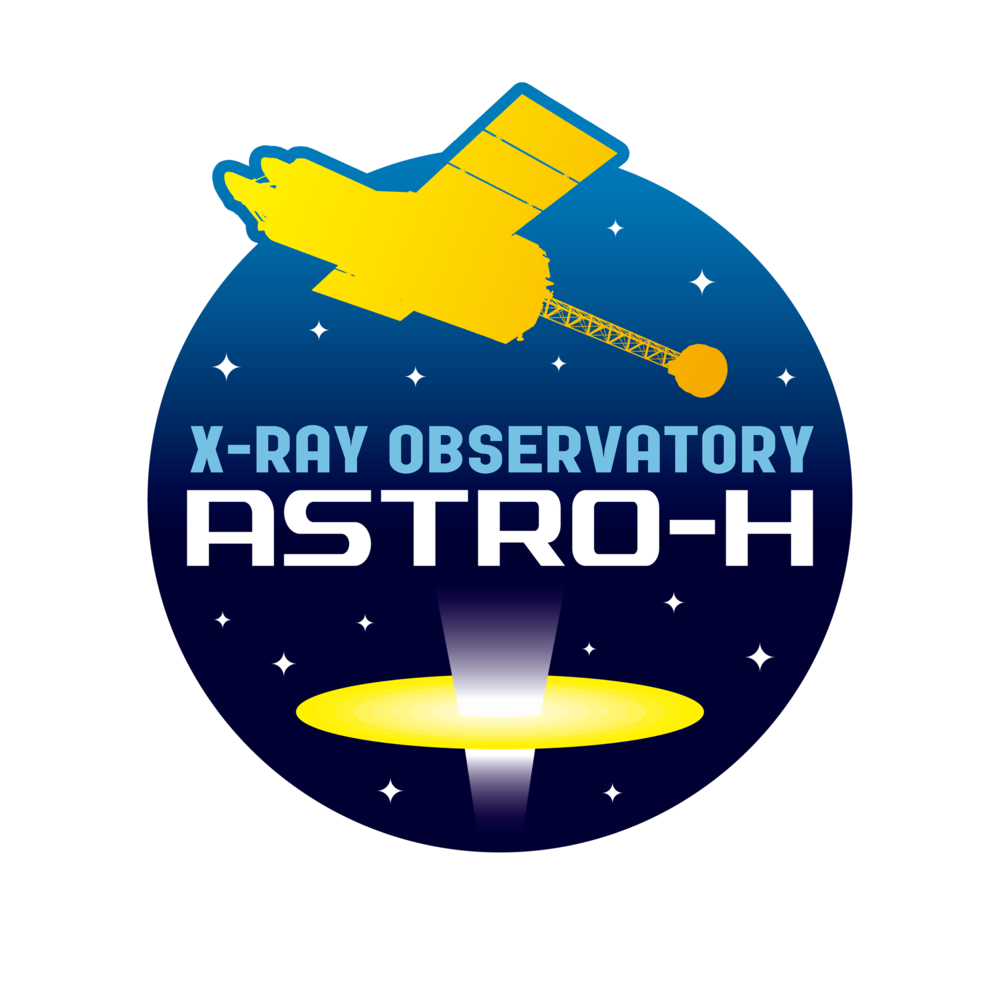}
			\end{center}
		\end{figure}
		\vspace{1cm}
		{\LARGE
		ASTRO-H Space X-ray Observatory\\
		White Paper\\
		}
		\vspace{5mm}
		{\large
		\WhitePaperTitle\\
		}
		\vspace{1cm}
		{
		\WhitePaperAuthors\\
		on behalf of the ASTRO-H Science Working Group
		}
	\end{center}
}

\usepackage{authblk}
\author[a]{Tadayuki~Takahashi}
\author[a]{Kazuhisa~Mitsuda}
\author[b]{Richard~Kelley}
\author[c]{Felix~Aharonian}
\author[d]{Hiroki~Akamatsu}
\author[e]{Fumie~Akimoto}
\author[f]{Steve~Allen}
\author[g]{Naohisa~Anabuki}
\author[b]{Lorella~Angelini}
\author[h]{Keith~Arnaud}
% ...
\affil[a]{Institute of Space and Astronautical Science (ISAS), Japan Aerospace Exploration Agency (JAXA), Kanagawa 252-5210, Japan}
\affil[b]{NASA/Goddard Space Flight Center, MD 20771, USA}
\affil[c]{Astronomy and Astrophysics Section, Dublin Institute for Advanced Studies, Dublin 2, Ireland}
\affil[d]{SRON Netherlands Institute for Space Research, Utrecht, The Netherlands}
\affil[e]{Department of Physics, Nagoya University, Aichi 338-8570, Japan}
\affil[f]{Kavli Institute for Particle Astrophysics and Cosmology, Stanford University, CA 94305, USA}
\affil[g]{Department of Earth and Space Science, Osaka University, Osaka 560-0043, Japan}
\affil[h]{Department of Astronomy, University of Maryland, MD 20742, USA}
% ...

\begin{document}

\newcommand{\WhitePaperTitle}{Low-mass X-ray Binaries}
\newcommand{\WhitePaperAuthors}{
	C.~Done~(Durham~University), M.~Tsujimoto~(JAXA),
	E.~Cackett~(University~of~Cambridge\footnote{Present address: Wayne~University}), J.~W.~den~Herder~(SRON), T.~Dotani~(JAXA), 
	T.~Enoto~(RIKEN), C.~Ferrigno~(Universit\`{e} de Gen\'{e}ve), T.~Kallman~(NASA/GSFC),
	T.~Kohmura~(Tokyo~University~of~Science), P.~Laurent~(CEA~Saclay),
	J.~Miller~(University~of~Michigan), S.~Mineshige~(Kyoto~University),
	H.~Mori~(Nagoya~University), K.~Nakazawa~(University~of~Tokyo),
	F.~Paerels~(Columbia~University), S.~Sakurai~(University~of~Tokyo), Y.~Soong~(NASA/GSFC),
	S.~Sugita~(Nagoya~University), H.~Takahashi~(Hiroshima~University), T.~Tamagawa~(RIKEN),
	Y.~Tanaka~(MPE), Y.~Terada~(Saitama~University), and S.~Uno~(Nihon~Fukushi~University)
}
\MakeWhitePaperTitle

\begin{abstract}
There is still 10-20\% uncertainty on the neutron star (NS)
  mass-radius relation. These uncertainties could be reduced by an
  order of magnitude through an unambiguous measure of $M/R$ from the
  surface redshift of a narrow line, greatly constraining the Equation
  of State for ultra-dense material. It is possible that the SXS on
  {\it ASTRO-H} can detect this from an accreting neutron star with low
  surface velocity in the line of sight i.e. either low inclination or
  low spin. Currently there is only one known low inclination LMXB,
  Ser X-1, and one known slow spin LMXB, J17480-2446 in Terzan 5. Ser
  X-1 is a persistent source which is always in the soft state (banana
  branch), where the accreting material should form a equatorial belt
  around the neutron star. A pole-on view should then allow the NS
  surface to be seen directly. A 100~ks observation should allow us to
  measure $M/R$ if there are any heavy elements in the photosphere at
  the poles. Conversely, J17480-2446 in Terzan 5 is a transient
  accretion powered millisecond pulsar, where the accreting material
  is collimated onto the magnetic pole in the hard (island) state
  ($L_x < 0.1 L_{Edd}$). The hotspot where the shock illuminates the
  NS surface is clearly seen in this state.  A 100~ks ToO observation of
  this (or any other similarly slow spin system) in this state, may
  again allow the surface redshift to be directly measured. 
 
NS LMXB can also show winds from the outer disk, detected as H
  and He-like iron K$\alpha$ absorption lines. The most likely origin
  of these winds is from thermal driving (helped by radiation pressure
  when $L\sim L_{Edd}$). This makes clear predictions about the launch
  radius, acceleration, and mass loss rate in the wind. Each of these
  can be constrained with a $50-100$~ks observation of the wind in the
  persistent source GX13+1. The ratio of emission to absorption (P
  Cygni profile) gives the solid angle of the wind, required to
  determine mass loss rates, while resolving the velocity width of the
  absorption gives much better constraints on the launch radius than
  derived from CCD data, and the detailed profile of the blue wing of
  the absorption traces the acceleration. This will critically test
  the thermal wind model of the disk wind.
\end{abstract}

\maketitle
\clearpage

\tableofcontents
\clearpage

\section{Introduction}

Low Mass X-ray Binaries (LMXB) were the first compact accretion
powered sources to be discovered (Sco X-1), where a neutron star (NS)
accretes via Roche Lobe overflow from a companion star. NS have
similar $M/R$ to the last stable orbit around a moderate spin black
hole (BH), so their accretion flows should be similar. The difference
is that the NS have a solid surface, while black holes do not. The
surface means that there is the possibility of X-ray bursts from
nuclear burning of the accreted material onto the surface, or coherent
pulsations from magnetically collimated accretion as well as a
boundary layer between the flow and the surface which produces both an
additional emission component and an additional source of turbulence
(high frequency noise). However, the most important possibility from
the surface is that there could be atomic features. Their
gravitational redshift would give a direct, accurate, measure of $M/R$
to observationally constrain the quantum chromodynamics (via the
equation of state) at densities which are far beyond those currently
accessible to laboratory experiments.

Another difference between the NS and BH is that the
NS has smaller mass. Thus a given type of companion star
must be closer in order to overflow its Roche lobe. The disk is
truncated by tidal forces at about half the orbital separation so the
disk is smaller and hence less likely to dip below the Hydrogen
ionisation temperature which causes the disk instability (King et al.
1996). Thus many NS are persistent sources, making targeting and
scheduling easier than in the typically transient BH.  Disk-jet
coupling, the physics of the accretion flow, the origin of the
quasi-periodic oscillations (both low and high frequencies), and the
origin of winds from the disk are all issues which can be addressed
using the NS systems as well as the BH. For example, disk accreting NS
LMXB (atolls) show the same hard/soft spectral transition as seen in
BH, going from the island state (hard) to banana branch (soft) to Z
sources (Eddington and above) as the mass accretion rate increases
(Lin, Remillard \& Homan 2009). 

The disk size is also important in determining whether the system
should power an equatorial disk wind. X-ray illumination heats the
upper layers of the disk to the Compton temperature. At small radii
this is still bound, so forms a static `corona' (a confusing name as
this in unrelated to the intrinsic hard X-ray source which is also
termed `coronal'). At larger radii, the velocity associated
with the Compton temperature is sufficient for the material to escape
as a thermally driven wind. Since NS typically have smaller disks than
BH, winds should be rarer, seen only in the largest separation binary
systems. 

The vertical structure from the static `corona' (and wind), together
with the interaction of the accretion stream with the disk, gives an
inclination dependence to LMXB spectra. At low inclinations there is
an unobscured view of the inner disk, at higher inclinations there can
be absorption lines from the `corona' (and wind), at even higher
inclinations there are also periodic absorption dips from the clumpy
impact of the stream and disk (dippers), and at extreme inclinations,
we see the central source only via scattering in the `corona' as a
direct view is obscured by the disk and/or companion star (Accretion
Disk Corona: ADC sources e.g. Frank, King \& Raine 1987).

However, the key property of the NS, that of its surface, gives rise
to some distinctive behaviour. LMXB are old systems, and the NS
surface field is low (below $10^{10}$~G in order to show bursts).
Measured fields are a few~$\times 10^{8}$~G from the handful of
systems which show magnetically collimated accretion - the accretion powered
millisecond X-ray pulsars (AMXP). Surface fields must be lower in the
transient NS as these do not show pulses at similarly low mass
accretion rates. This is generally explained by them having a higher long
term average mass accretion rate which buries a similarly strong
surface field (Cumming et al. 2001, but see Lamb et al. 2009 for an
alternative).

Unlike the BH in LMXB (King \& Kolb 1999), the spins of NS in LMXB
have changed substantially during their lifetime. Firstly, they spin
down from their birth via a rotation powered pulsar phase, then when
the binary makes contact they spin up via the accretion of angular
momentum as seen by the fast spins inferred from burst oscillations in
LMXB (see e.g. the review by Patruno \& Watts 2012). The accretion
rate declines as the binary evolves, so the magnetic field can emerge
from the surface and give an AMSP. Eventually the accretion rate
becomes so low that the NS can re-emerge as a low field ($\sim
10^8$~G) rotation powered millisecond pulsar, where the pulsar wind
ablates the companion star, hastening its demise (black widow pulsars
and red backs: Papitto et al. 2013).

\section{The Equation of State for Nuclear Matter}

\subsection{Background and Previous Studies}

The mean density of a NS is $2-3\times\mathrm{nuclear}$, and their central densities
are much higher. This is a regime in which quantum chromodynamics
cannot be tested in laboratory experiments, so there is considerable
uncertainty in the equation of state (EoS) models.  Current
constraints from mass measurements of NS already favour a
'normal nucleon' EoS, giving predicted radii which do not depend much
on mass from $1-2M_\odot$ at $11-12$~km ($\sim 7-3.5R_g$ e.g. Ozel
2013). However, there are still a range of models within this region,
and substantial model uncertainties even for a given EoS (Hebeler et
al. 2013). Better astrophysical data would give clear input to the
nuclear physics community to constrain their models.

Constraints on $M/R$ can be made in a variety of ways:

a) redshift of any narrow lines emitted from the photosphere of the
NS (SXS)

b) redshift of the broad iron line from reflection from the accretion
disk as the inner edge of the disk is an upper limit on the size of
the NS (SXS to get detailed line shape without pileup and
resolve out narrow emission/absorption features, plus
broad band spectrum to constrain the complex continuum)

c) detailed pulse profile modelling of accreting millisecond pulsars
and burst oscillations (broad band spectra)

d) a direct measure of the radius from the luminosity and temperature
of thermal radiation from the NS surface cooling after accretion has
stopped (broad band spectrum in quiescence)

The problem is important enough that all of these approaches need to
be used as it is only when we get consistent answers from multiple
methods that there can be real confidence in the result. Nonetheless,
only measuring redshift from a narrow line offers the possibility of
significantly improved constraints on $M/R$ with {\it ASTRO-H}. 

\subsection{Prospects \& Strategy}

The SXS effective area around the iron K alpha line is almost an order
of magnitude larger than the {\it Chandra} HETG, the only other high
resolution instrument to cover this transition (225 versus 28~cm$^2$
at $6.5-7$~keV). However, this effective area gain in {\it ASTRO-H} is offset
by the maximum count rate limit for pileup in bright sources, with the
two instruments having equal usable counts at iron for a flux of $\sim
4\times 10^{-8}$~ergs~cm$^2$~s$^{-1}$ for a soft transient with
typical Galactic $N_H\sim 6\times 10^{21}$~cm$^{2}$. Nonetheless, weak
and {\it narrow} features are still more easily detected and
characterised in {\it ASTRO-H} as its resolution is better by a factor $\sim
8-10$ at iron, giving it significant discovery space even in bright
sources.

To see narrow absorption lines from the NS surface requires
that there are heavy elements in the photosphere, that these are not
completely ionised, that the photosphere is not buried beneath the
accretion flow, and that the resulting atomic features are not
substantially broadened by thermal, magnetic or rotational effects.

Heavy elements are deposited onto the NS surface by the accretion
flow. They are stopped by collisional processes, which are more
efficient for higher mass/charge ions. Hence iron and other heavy
elements are halted higher up in the photosphere than lower atomic
number elements. They can then be destroyed by spallation bombardment
(by the still energetic helium and hydrogen ions, transforming the
iron nuclei to lower Z elements) or sink under gravity.  The
deposition and destruction rate both depend linearly on $\dot{M}$ so
the steady state Fe column is around solar, independent of $\dot{M}$
for $L_x>6\times 10^{32}$~ergs/s (Bildsten, Chang \& Paerels 2003;
Chang, Bildsten \& Wasserman 2005). Thus there can be iron in the
photosphere of an accreting NS, especially as the outwards radiation
force on the line transitions act against gravity to keep iron (and
other elements) from settling where these are not completely ionised
(Chang, Bildsten \& Wasserman 2005) i.e. for temperatures of less than
$\sim 1.5$~keV (Suleimanov, Poutanen \& Werner 2012). 

\begin{figure}[t] 
\begin{center} 
\begin{tabular}{llll}
\includegraphics[scale=0.14]{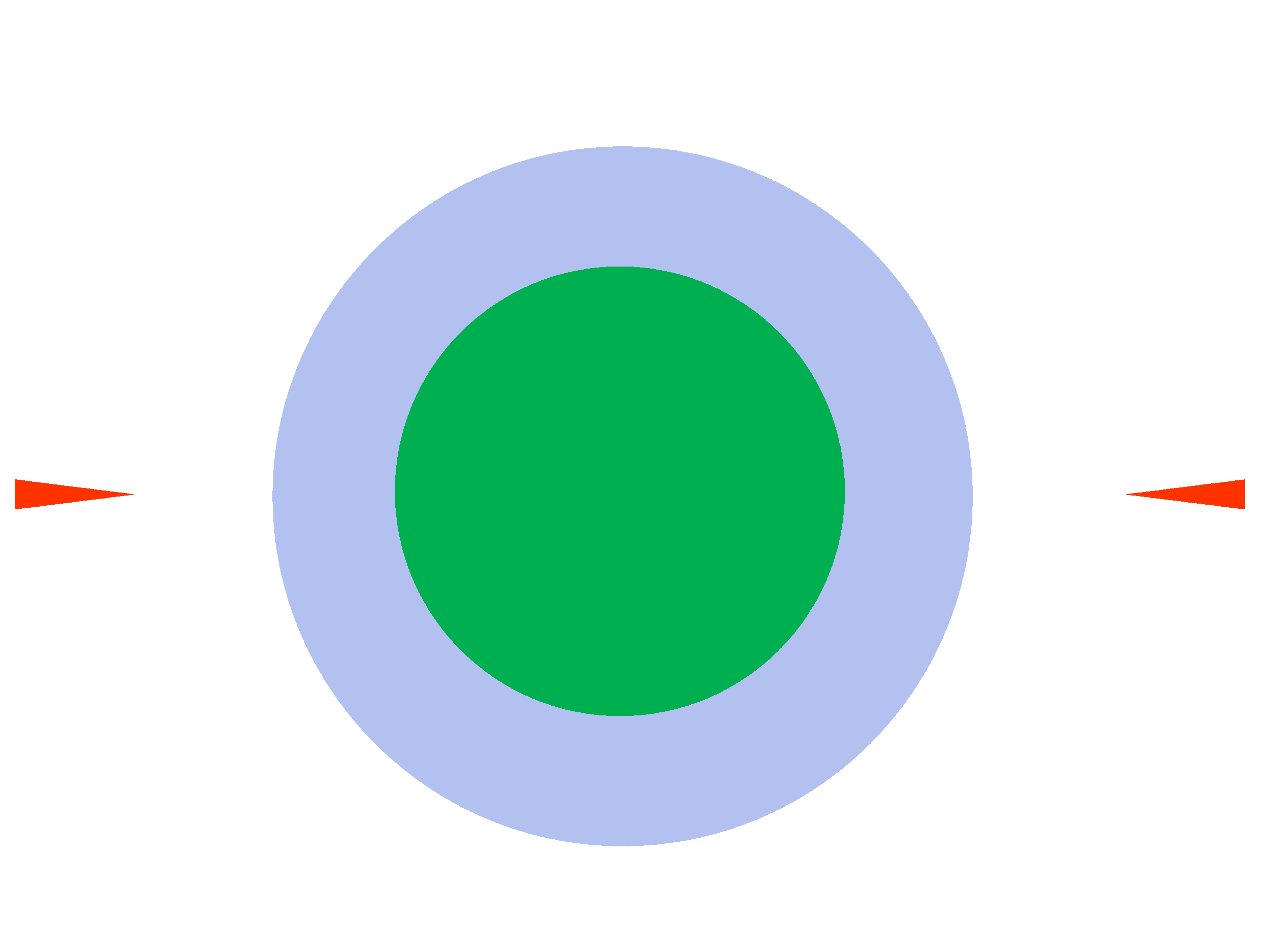} &
\includegraphics[scale=0.14]{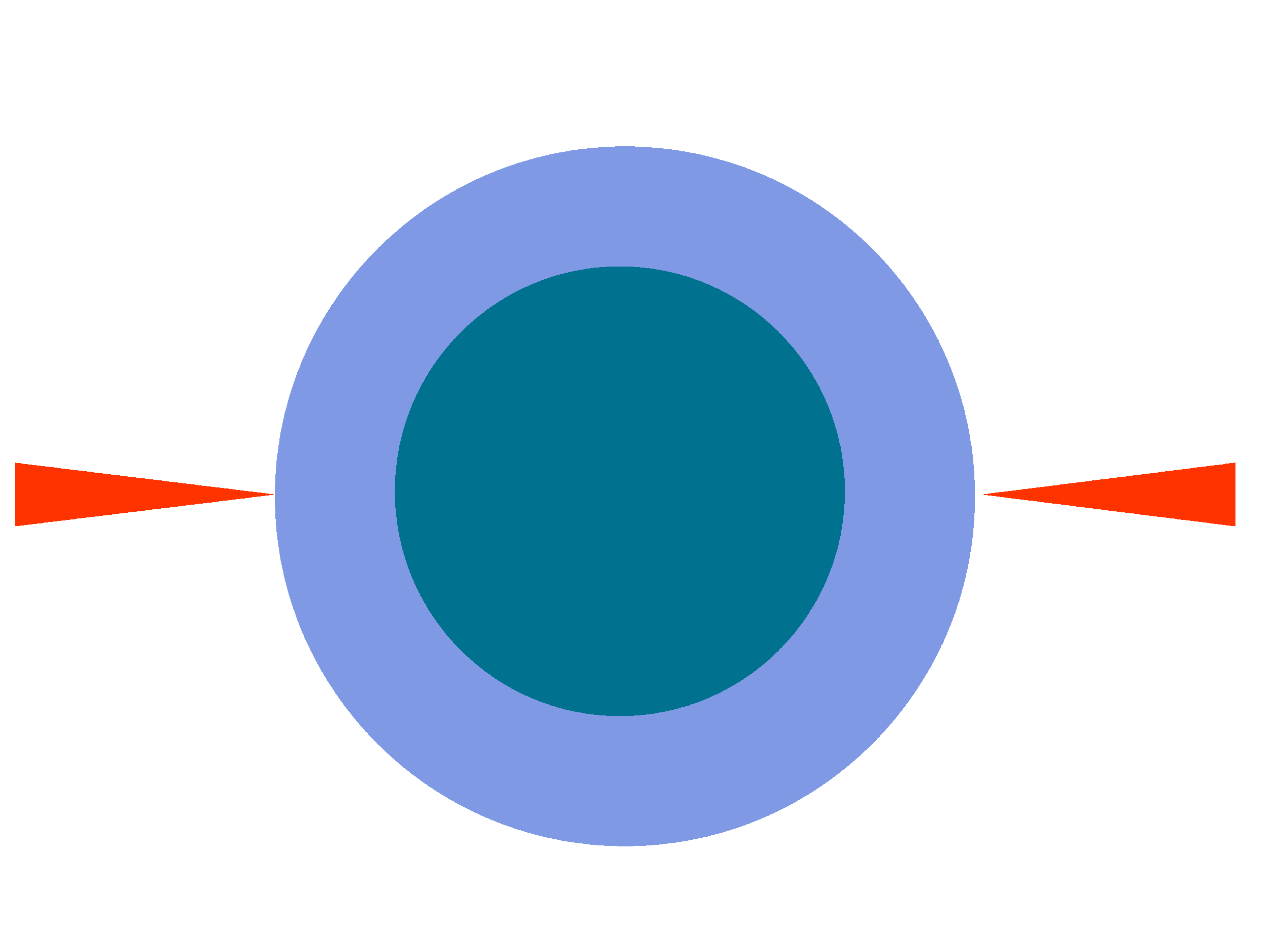} &
\includegraphics[scale=0.14]{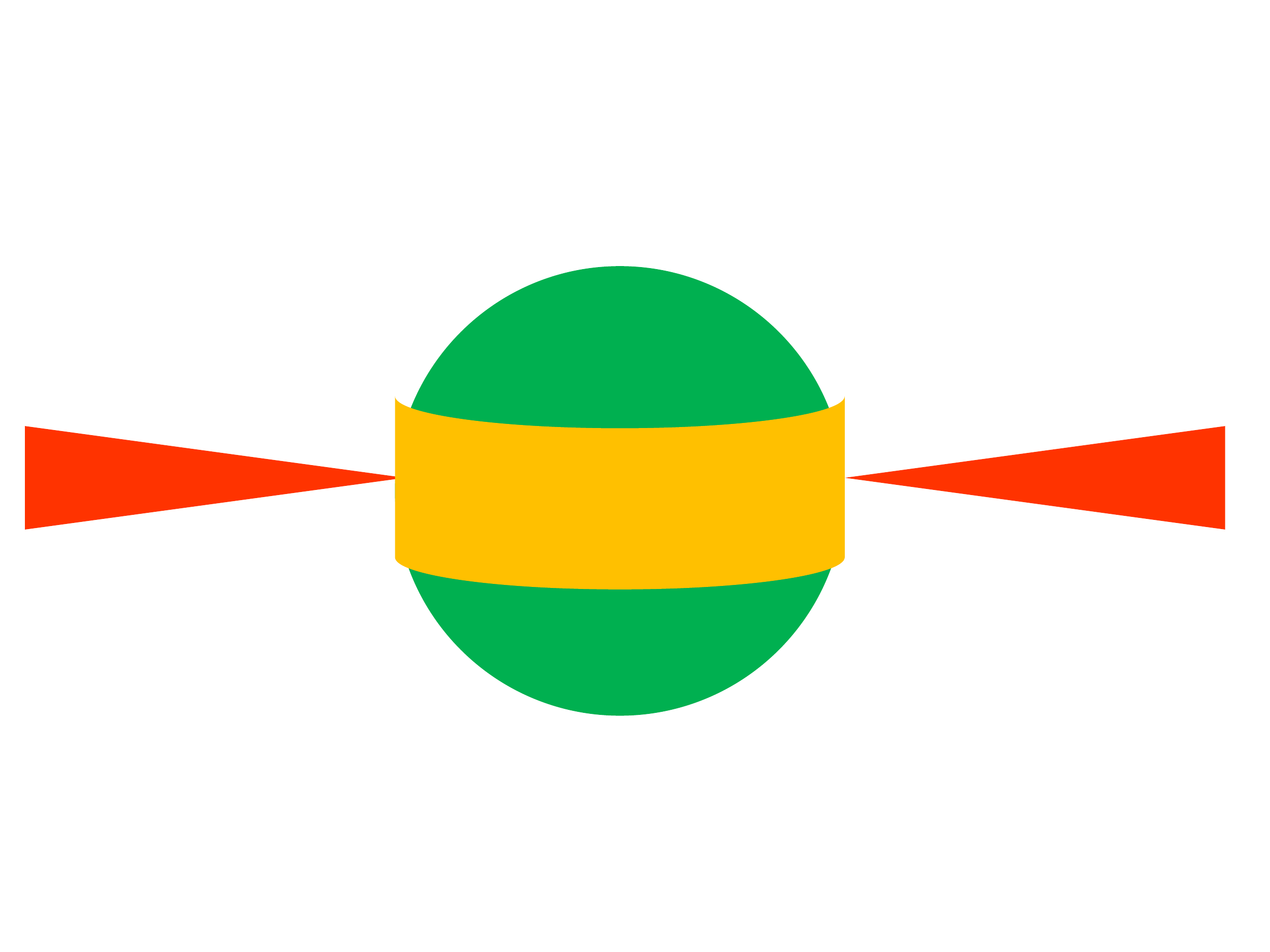} &
\includegraphics[scale=0.14]{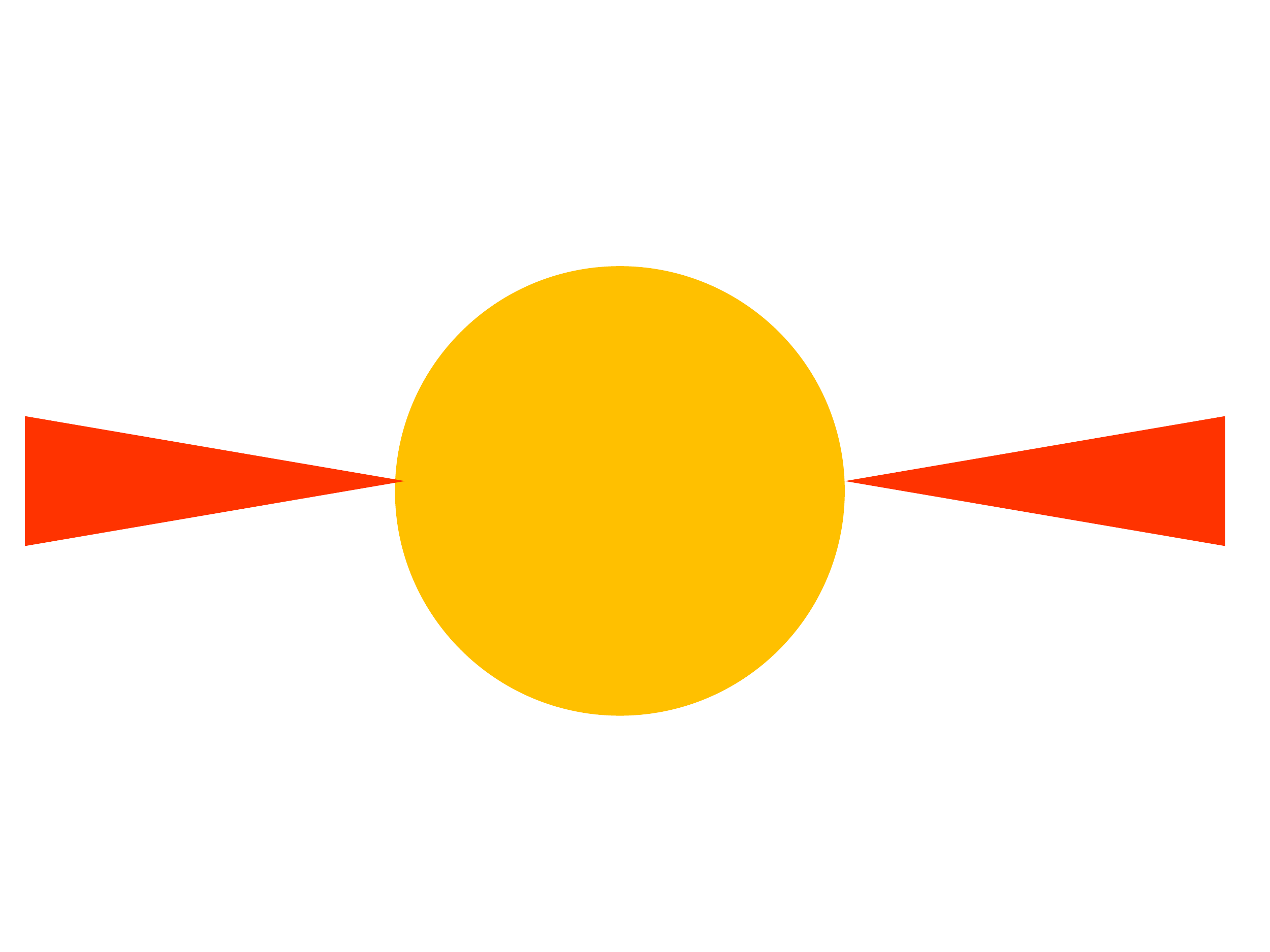} 
\end{tabular}
\end{center}
\vspace{-32pt}
\caption{From left to right shows the changing geometry of the
  accretion flow onto the NS surface with $L/L_{Edd}$ increasing from
  $\sim 10^{-4}$, $\sim 10^{-2}$, $\sim 0.1$ and $\sim 1$. At the
  lowest mass accretion rates the disc (red) is truncated far from
  the NS surface (green), and the inner accretion flow/boundary layer
  is hot and optically thin (blue). The flow 
  becomes more optically thick but remains hot as the mass accretion
  rate increases in the island state, so that the neutron star surface
  can still be seen through the translucent flow. The flow then 
  collapses into a thin disc, so the boundary layer (yellow) forms an
  optically thick equatorial belt, which then extends upwards with
  increasing mass accretion rate to cover the entire surface. 
}
\label{fig:nsgeom} 
\end{figure}

The surface can only be seen if it is not hidden beneath the accretion
flow. This depends on the geometry of the accreting material as well
as its optical depth. At low accretion rates, the accretion flow is
hot and quasi-spherical interior to some truncation radius at which
the thin disk evaporates (island state). There is additional
luminosity from the boundary layer where this settles onto the
surface, but the flow and boundary layer merge together, forming a
single hot ($\sim 30-50$~keV), optically thin(ish) ($\tau\sim 1.5-2$)
structure (Medvedev \& Narayan 2001).  A fraction $e^{-\tau}$ of the
surface emission should escape without scattering, so $\sim 10-20$\%
of the intrinsic NS photosphere should be seen directly. The
temperature of this surface emission can also be seen imprinted onto
the low energy rollover of the Compton spectrum and is only $\sim
0.5-0.6$~keV (e.g. Sakurai et al. 2012), so there should be atomic
features (see Figure~\ref{fig:nsgeom}a and b).

At higher mass accretion rates ($L\ge 0.1-0.5L_{Edd}$), the thin disk
extends down to the NS surface, forming a boundary layer where it
impacts around the NS equator. The boundary layer is now optically
thick ($\tau\sim 5-10$) so hides the surface beneath it, though the
temperature can still be seen via its imprint on the low energy
rollover of the Compton spectrum at $\sim 0.8-1.3$~keV. The boundary
layer itself is at the local Eddington temperature of $\sim 2.5$~keV
(Revnivtsev, Suleimanov \& Poutanen 2013), so is completely
ionised. However, the vertical extent of the boundary layer depends on
the accretion rate, and it only covers the entire surface, extending
up the NS pole only when the accretion rate is around Eddington
(Suleimanov \& Poutanen 2006, Figure~\ref{fig:nsgeom}c and d).

Thus the surface is always at low enough temperatures to not be
completely ionised, but is completely covered by the optically thick,
completely ionised accretion flow at $L\sim L_{Edd}$ (upper banana
branch and Z sources).  It is also completely covered by the much
hotter (so still completely ionised) accretion flow at low mass
accretion rates (island state), but here the material is not so
optically thick so a fraction of the surface can be seen directly.
Hence the largest fraction of surface emission would be seen from a
pole-on view of a lower banana branch source, where the optically
thick accretion flow is confined to an equatorial belt, or in an
island state, where the accretion flow covers most of the surface but
is optically thin(ish).

The surface can also be seen directly during X-ray bursts. Bursts have
the advantage that there are almost certainly some heavy elements
present in the photosphere due to dredge up in the thermonuclear
explosion.  However, the bursts reach high enough temperatures that
all elements are completely ionised (Suleimanov et al. 2012). 
Bursts occur over only a small fraction of a total observation,
and selecting only the cooler phases of the burst will limit the
signal-to-noise even further, making this unlikely to be feasible.

Looking at the surface emission outside of the bursts may then be the
better strategy. Any atomic features will be redshifted to $E$ from
$E_0$ depending on the radius of the star $r=R/R_g\propto R/M$ where
$r=2/[1-(E/E_0)^2]$ and $E/E_0\sim 0.8$ for typical NS parameters
($r=6$ corresponds to 12.6~km for $1.4M_\odot$ i.e. a surface redshift
of 0.22, while a radius of 10~km gives a redshift of 0.30).  This
shift is the same over the entire surface for a spherical star, with
thermal and Stark (pressure) broadening giving an intrinsic width of a
few eV. Magnetic fields in the line-forming region broaden the lines
via the Zeeman effect, giving $\Delta E\sim 12 B_9$~eV, where
$B_9=B/(10^9 {\rm G})$ (Loeb 2003), so as long as $B_9<1$ (as is usual
in LMXB) then this is still narrow for the SXS.

However, there is also rotational broadening, with $\Delta E
\sim 1600 (\nu_{spin}/600 {\rm Hz}) (R/10 {\rm km})$~eV (Ozel 2013),
and this is a clear issue as most LMXRB (i.e. systems with $B<10^9$~G
as required) have spins $\nu_{spin}=185-650$~Hz (Pappitto et al. 2011).

Thus the only way to have any significant chance to recover narrow
atomic features is to have an accreting NS where the surface is not
too highly ionised (i.e. not strong X-ray bursts) and can be seen
directly (i.e. not the high mass accretion rate upper banana branch or
Z sources).  More stringently, to see these features as narrow
requires that the line of sight velocity component from the surface is
small (slow spin NS, pole on faster spin NS, phase resolved small spot
on NS), and that the magnetic field is low.

Any target meeting these requirements should be a high priority PV
observation.

\subsection{Targets \& Feasibility}

\subsubsection{Low inclination}

Normal spin systems which are viewed at low inclination ($<30^\circ$)
can still show a narrow core to the line profile, even though the
majority of the line is rotationally broadened (Baubock, Psaltis \&
Ozel 2013). There is one atoll system, Ser X-1, where optical
spectroscopy of the 2 hour binary orbit indicates a low inclination,
$i<10^\circ$ (Cornelisse et al. 2013). A low inclination is also
consistent with the non-detection of dips in the X-ray lightcurves
(the structures in the MAXI long term light curves are not real:
Shidatsu, private communication), and the lack of any burst
oscillations in the X-ray burst lightcurves (Galloway et al.
2008). This persistent system is always in the soft state (mid banana
branch, $L/L_{Edd}\sim 0.3$ for 7.7~kpc), so the disk should
extend down close to the NS surface, consistent with the iron line
broadening seen by {\it NuSTAR} (Miller et al. 2013).

\begin{figure}[b] 
\begin{center} 
\begin{tabular}{lll}
\includegraphics[scale=0.25]{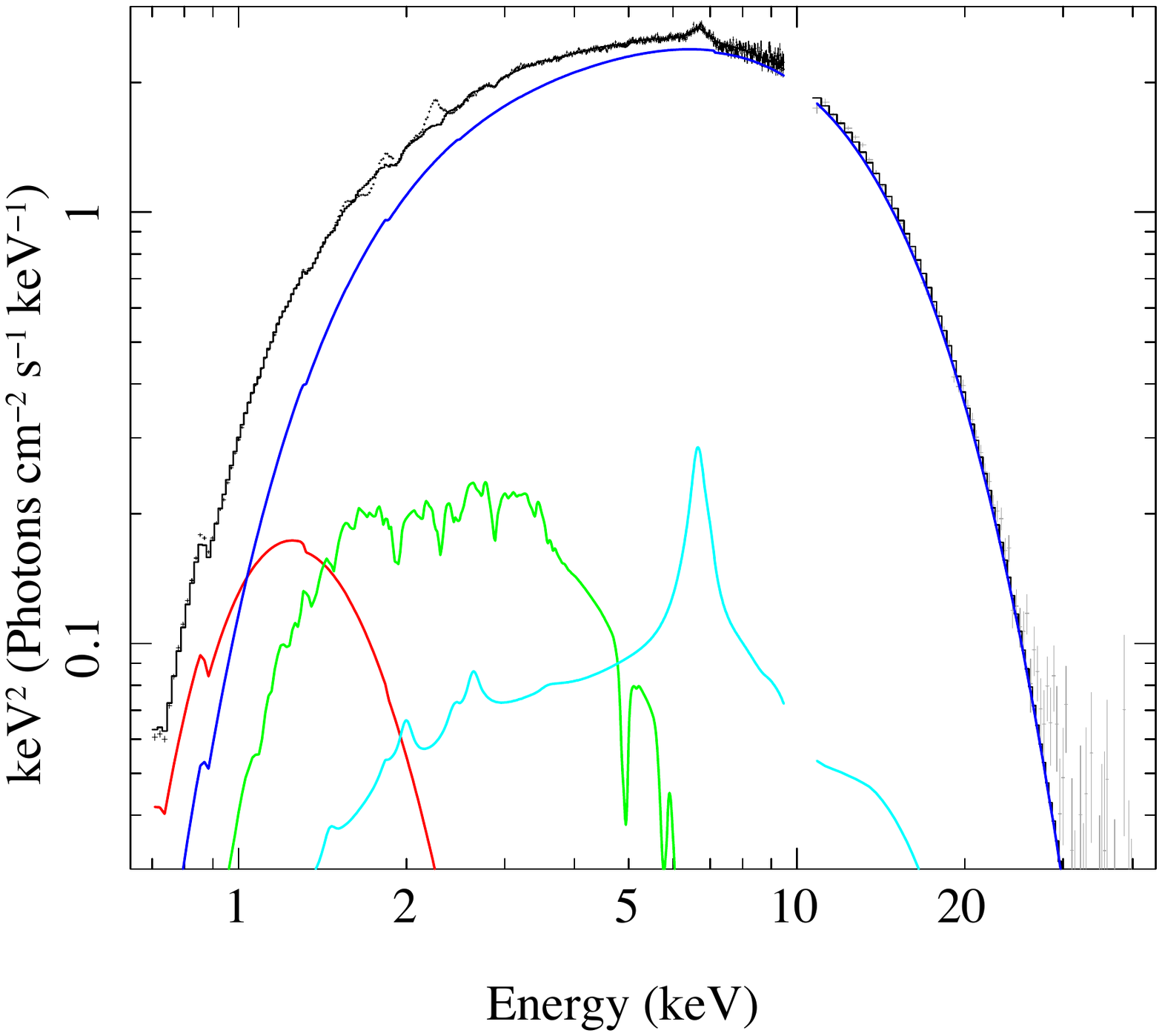} &
\includegraphics[scale=0.25]{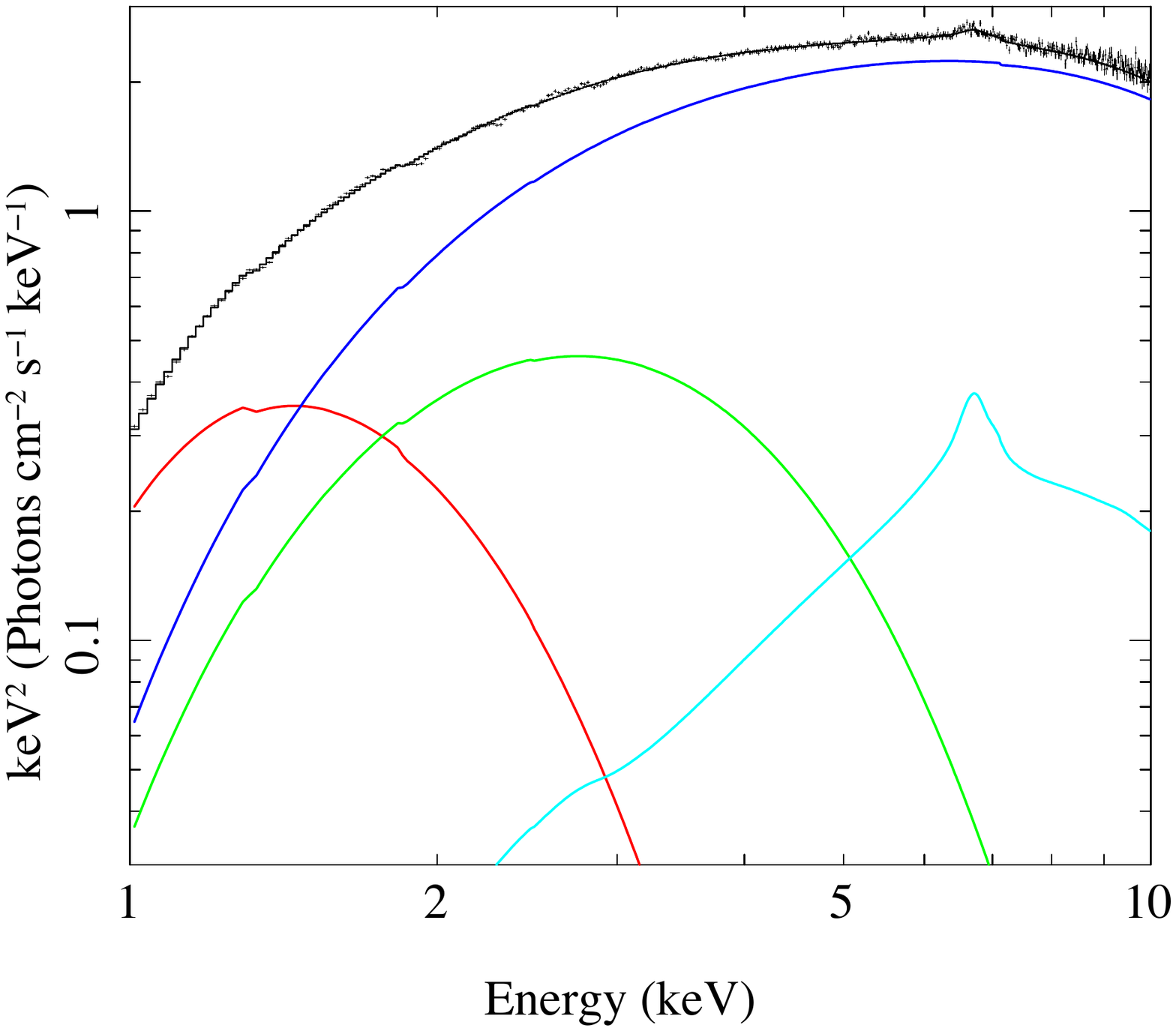} &
\includegraphics[scale=0.25]{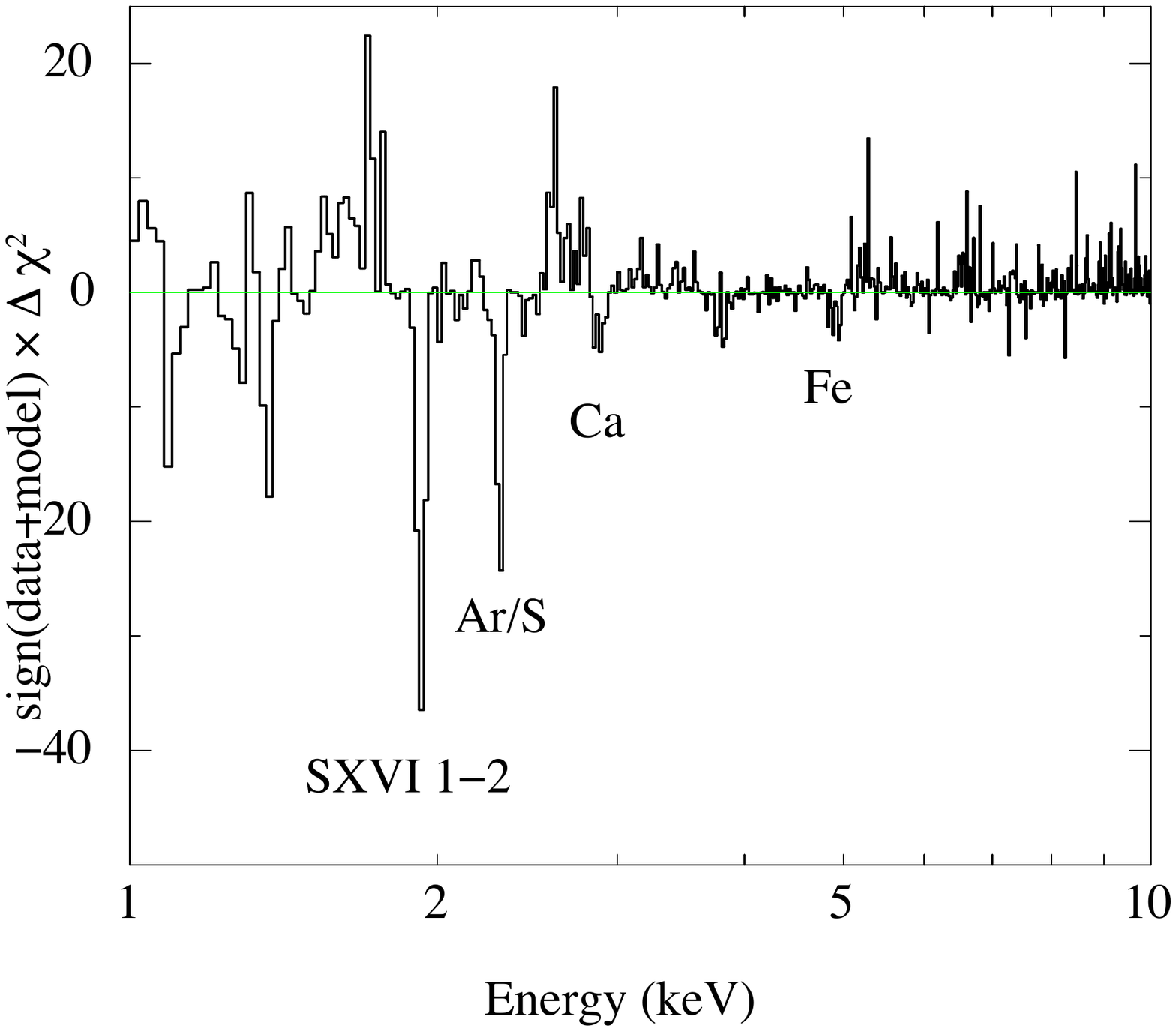}
\end{tabular}
\end{center}
\vspace{-32pt}
\caption{a) Ser X-1 {\it Suzaku} XIS (black) and HXD (grey) data, fit with 
  diskbb (red) + 6~M~K NS surface convolved with the transfer
  function expected for 10$^\circ$ inclination (green) + Comptonised
  boundary layer with seed photons at $\sim 6$~M~K  (blue) and its
  reflection (cyan). b) The spectrum in a) simulated through the SXS
  with the neutral density filter for 50~ks. The model is as in a)
  except that the NS surface is replaced by a blackbody (green). c)
  residuals from the data/model fit in b) clearly showing the 4 main
  absorption lines seen in the model in a). 
}
\label{fig1} 
\end{figure}

We fit the {\it Suzaku} XIS (black) and HXD (grey) data with a model
incorporating the disk (diskbb: red), NS surface (bb:
magenta), Comptonised boundary layer (nthcomp: blue) and its
reflection (kdblur$\times$rfxconv$\times$nthcomp: cyan). This gives
$\chi^2_\nu=1115/598$. The observed blackbody temperature from the NS
surface is $\sim 0.6$~keV. This is gravitationally redshifted, so the
intrinsic surface temperature is a factor $1.2-1.3\times$
higher. This intrinsic photospheric emission is not a true
blackbody, but can be approximated by a colour temperature corrected
blackbody, where the colour temperature correction factor $f_{col}\sim
1.4-1.5$ (Suleimanov et al. 2012). This cancels out most
of the redshift, so the intrinsic photosphere temperature should be
similar to the observed blackbody temperature at $\sim 0.6$~keV. This
is cool enough that there can easily be atomic features present.

We replace the blackbody in the fit with a NS photosphere model at
6M~K (V. Suleimanov, private communication). We convolve this with the
full transfer function expected from a NS rotating at 400~Hz (a
typical spin) viewed at $10^\circ$, assuming a mass of $1.4M_\odot$
and 10~km radius (M. Baubock, private communication), corresponding to
$\log g=14.27$. This is effectively the same as convolving the
photosphere model with {\sc gsmooth} with $\Delta E=0.05$~keV at
6~keV, and $\Delta E/E=1$, together with a redshift
$z_\mathrm{surface}=0.305$.  This shows that there are 4 main absorption
features at this temperature (see Fig 1a), FeXXV at 1.8508\AA, a blend
of ArXVII Ly$\alpha$ (3.9493\AA) with SXVI Ly$\beta$ (3.9908\AA) at
mean wavelength of 3.9700\AA, CaXIX at 3.1773\AA and SXVI at
4.7328\AA.

We fit the full photosphere/transfer function model to the {\it Suzaku}
data, but allow the NS temperature to be slightly different
than the tabulated model by multipling the 6M~K template by a further
redshift ({\sc zashift}=0.039), showing that the data prefer a
slightly lower intrinsic photosphere temperature. We get a slightly
better fit for this more realistic photosphere model than with the
pure blackbody, with $\chi^2=997/597$ (one fewer degree of freedom as
the seed photons for Comptonisation are no longer tied to the NS temperature) but this is due to the fact that the data prefer the
broader continuum over a narrower blackbody, rather than {\it Suzaku}
detecting the lines. The 4 main lines have equivalent width of 2.2eV
(FeXXV), 1.8~eV (CaXIX), 1.8eV (ArXVII/SXVI) and 1.9eV (SXVI) against
this continuum model. 

We simulate this model, which includes the photospheric lines with
total redshift $z_\mathrm{surface}=1.305\times1.039-1=0.356$, through the SXS,
using the 7eV resolution response. This gives a count rate of $\sim
350$ c/s so we use the neutral density filter to reduce this to 88
c/s. We resimulate this for 50~ks and 100~ks, and fit each dataset in
the 1-10~keV bandpass with the original {\it Suzaku} model i.e. with a blackbody
for the NS surface.  The lines are clearly detected in all
simulations by plotting the resulting $\chi^2$.

We then add in the 4 narrow gaussians, all with the same broadening
({\sc gsmooth} with $\Delta E=0.05$~keV at 6~keV, and $\Delta E/E=1$)
and redshift ({\sc zashift}). These are very significantly detected in
both simulations, with $\Delta\chi^2=136$ and $220$ for 50~ks and
100~ks, respectively. Both return the input $z_\mathrm{surface}=0.356$ both
with uncertainties of $\sim 1$\% due to the width of the lines. 

The significance of the detection is mainly due to the two low energy
lines, as can clearly be seen in the $\chi^2$ residual plot to the
continuum models. In the 50~ks simulation, $\Delta\chi^2$ for each
line is $15$ (Fe), $22$ (Ca), $53$ (Ar/S) and $47$ (S).  Thus this is
clearly feasible to detect all 4 lines even in 50ks, but this did not
allow for pileup/PSP/telemetry limitations.  To derive a similarly
good spectrum in only high+medium resolution counts will probably take
100ks of observing time.

However, the most significant lines are at energies where it should
also be possible for the {\it Chandra} gratings to detect them, though chip
gaps mean the sensitivity is reduced at 1.95 and 2.3~keV.  There are
currently 150ks of {\it Chandra} METG/HETG data (soon to be increased by
another 150ks). A preliminary analysis of these data (E. Cackett,
private communication) shows no obvious features at this level.  A
more detailed analysis of the {\it Chandra} grating data should be able to
constrain the line equivalent widths even if they are a factor 2
smaller than predicted here. An upper limit (as opposed to a marginal
detection) from {\it Chandra} would imply the photosphere abundances of
these elements at the poles is less than $\approx$
$0.5-1\times$~solar.

The models used here assumed solar abundances, as indicated from the
strength of reflected iron line observed from the accretion
disk. There is significant uncertainty on the abundance
distribution over the NS surface both due to the spallation (Chang et
al. 2005), and due to the spatial inhomogeneity of the accretion flow.
Accretion is expected to occur in an equatorial belt, so a pole-on
view gives an unobscured view of the NS surface, but could be 
dominated by material where the heavy elements have settled out and
are not replenished by accretion. This source has X-ray bursts
(Galloway et al. 2008, 0.1 per hour so expect 2-3 in 100~ks, though
there appear to be none in the 150ks {\it Chandra} data) which should dredge
up the elements over the whole surface, but these can sink rapidly
after the burst (see e.g. Ozel 2013), A 100ks {\it ASTRO-H} observation
would tightly constrain the elemental composition of the NS crust at
the poles.

The cool phase of the bursts can additionally be searched for atomic
features since the abundances here may be extremely supersolar due to
dredge up.

\subsubsection{Beyond Feasibility: low inclination}

All the burst spectra can give 10-20\% constraints on $M/R$ from the
measured luminosity and temperature of the burst. 

The HXI and SGD together will trace the shape of the continuum at high
energies. The {\it Swift} BAT already shows that there is no persistent hard
tail which extends out to higher energies, but the {\it Suzaku} HXD and
{\it NuSTAR} data show that the shape of the rollover is not completely
matched by a single temperature Comptonised spectrum and its
reflection. Simultaneous data at high energies will show whether the
Comptonised boundary layer has some range in temperature, as predicted
by models of the spreading equatorial belt (Suleimanov \& Poutanen 2006).

The reflection spectrum from the accretion disk again can give 10-20\%
constraints on $M/R$ from the iron line profile (e.g. Miller et al. 2013).

\subsubsection{Slow spin}

The slowest known NS spin system is IGR J17480-2446 in Terzan 5 (T5X2
hereafter) where $\nu_{spin}=11$~Hz, equivalent to 30~eV in rotational
broadening.  However, this is also an AMXP, with inferred magnetic
field of few~$\times 10^{8}-10^{10}$~G (Miller et al. 2011; Cavecchi
et al. 2011; Patruno \& Watts 2012). Zeeman broadening dominates over
rotational broadening at the upper end of this range, with predicted
width of 100~eV, but the detected high frequency QPO in this system
constrains the field to the lower range (Barret 2012). 

The next slowest spin system is the accreting millisecond pulsar XTE
J0929-314 at 185~Hz (the 45~Hz claimed spin from EXO0748 is now known
to be spurious with the true spin at 550~Hz, and the 95~Hz signal from
XB 1254-690 is only a 2$\sigma$ detection see e.g. the compilations in
Patruno \& Watts 2012; Ozel 2013). Hence rotational broadening is over
300~eV, making the features undetectable.  Thus currently the only
known slow spin system is T5X2. 

As this is an AMXP the accretion geometry can be rather different to
that of the standard LMXB. At low mass accretion rates, the
quasi-spherical hot accretion flow can be collimated by the B field,
forming a shock when it hits the NS surface. Hard X-ray emission from
the shock illuminates the NS surface close to the magnetic pole,
forming a hotspot. Evidence for this is seen from the different pulse
profiles of the (blackbody) hotspot and hard X-ray Comptonised
emission from the shock (e.g. Gierlinski, Done \& Barret 2002). Hence it is
clear in this state that we do see part of the NS surface directly.
There should also be a cooler component from the unilluminated NS
surface further from the shock, and an even cooler component from the
truncated thin disk (e.g. Kajeva et al. 2011).

\begin{figure}[t] 
\begin{center} 
\begin{tabular}{llll}
\includegraphics[scale=0.14]{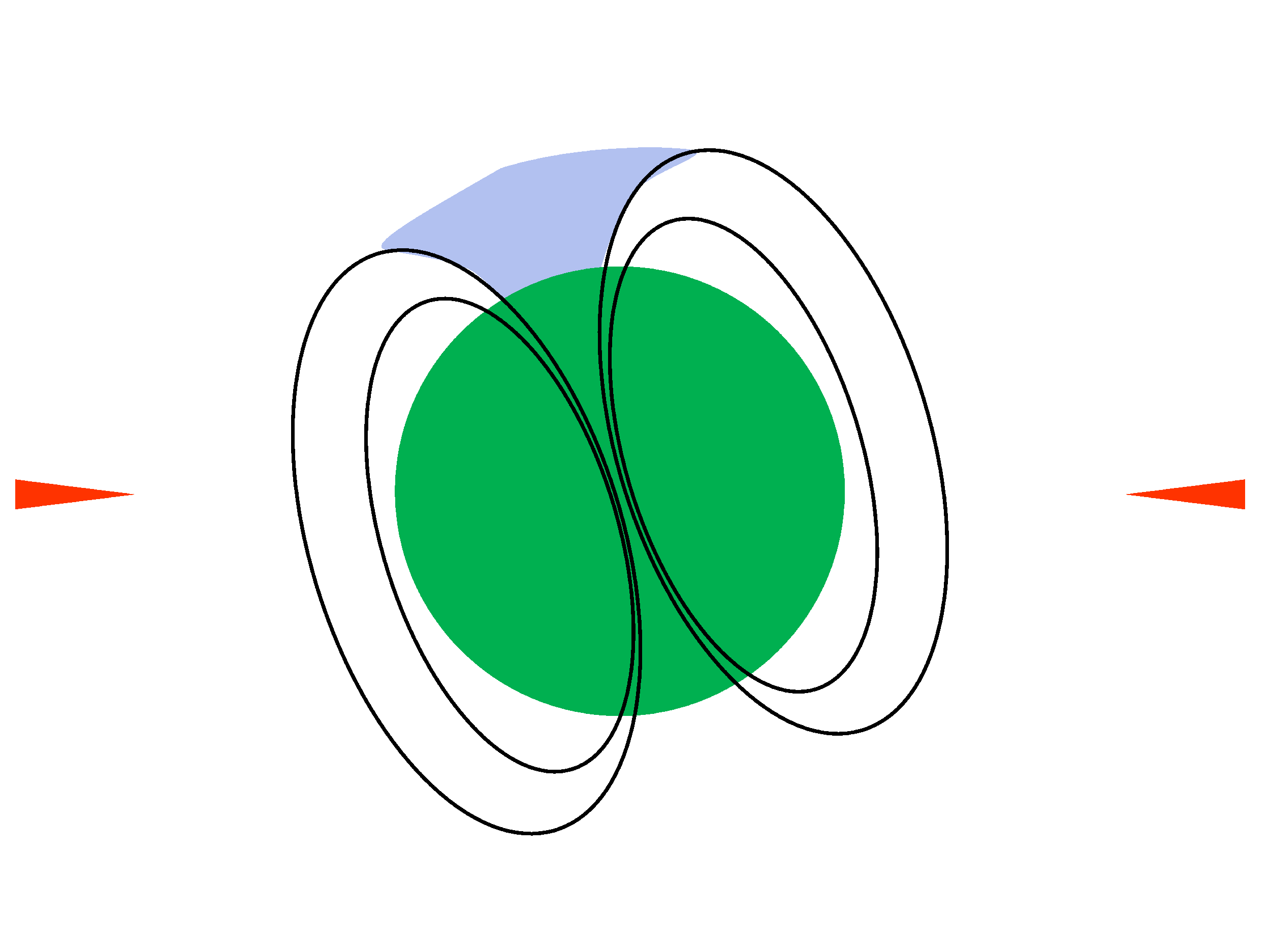} &
\includegraphics[scale=0.14]{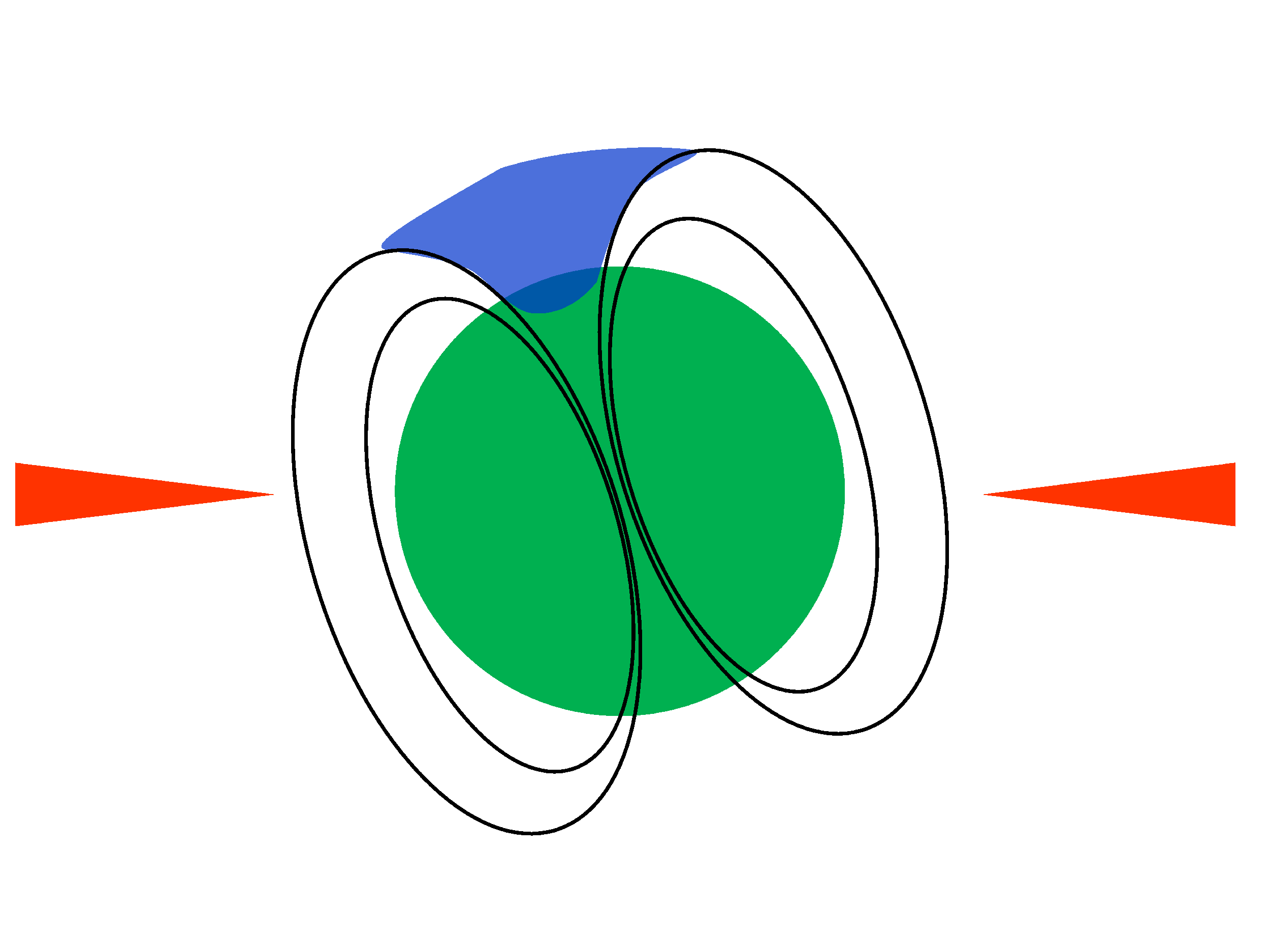} &
\includegraphics[scale=0.14]{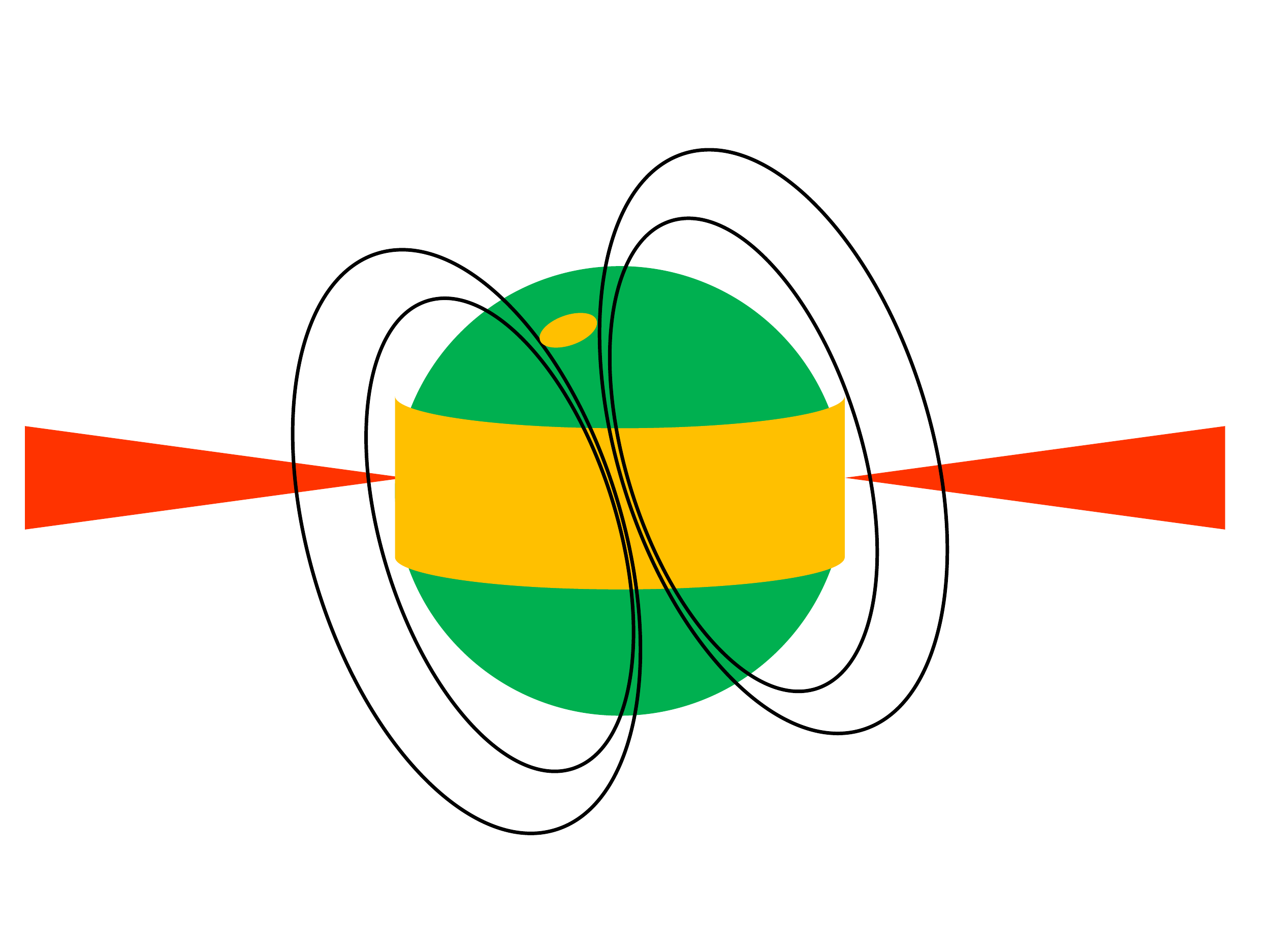} &
\includegraphics[scale=0.14]{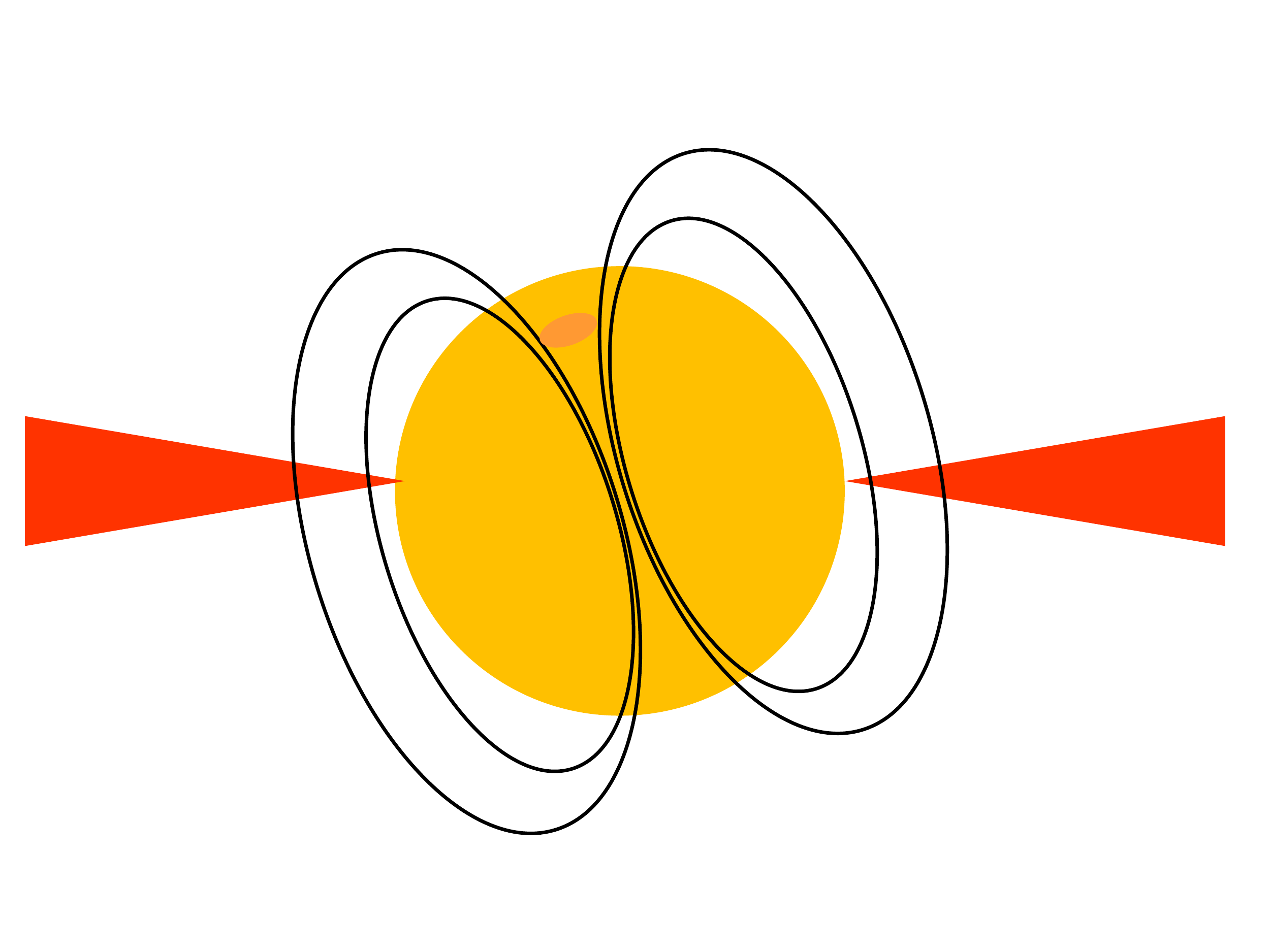} 
\end{tabular}
\end{center}
\vspace{-32pt}
\caption{As in Figure~\ref{fig:nsgeom} but for the AMXP where the
  magnetic field (black) is dynamically important and can collimate
  part of the accretion flow onto the magnetic poles. 
}
\label{fig:nsgeomb} 
\end{figure}

T5X2 is also the only known AMXP to go to high enough mass accretion
rates to go into the soft state. Here the disk should extend down
(close to) the NS surface, and the pulsed fraction drops abruptly
(though it is still detectable: Papitto et al. 2012). In fact, this
source goes to such high luminosities that it becomes a Z source
(Altamirano et al. 2010). There is a {\it Chandra} grating observation 5 days
after the Z source peak, when the source had already dimmed down to
the banana branch ($L/L_{Edd}\sim 0.3$ for 5.9~kpc: Papitto et al.
2012; Chakraborty et al. 2011). The continuum spectrum is very similar
to that from Ser X-1, and again there are no features from the NS
surface in the {\it Chandra} data (Miller et al. 2011), though here the NS
surface may be less visible than in Ser X-1 as T5X2 is not pole on.
Thus for this source there is more chance of detecting the surface
in an island state rather than on the banana branch. 

T5X2 is a transient, and is currently in quiescence so we need to wait
for another outburst. Figure~\ref{fig1}a shows a model for a bright
island state of SAX J1808.4-3658 (hereafter J1808) from Kajava et al.
(2011), where the total (black) includes the disk (red), neutron star
surface (green) and comptonised emission (blue). The grey line shows
the NS surface from a 6M~K solar abundance thermal NS model
(V. Suleimanov, private communication), with surface redshift of 0.3.
This is broader than than the blackbody, but this roughly matches the
peak energy and normalisation component seen in the data.

We convolve the NS surface model with {\tt gsmooth} with $\Delta
E/E=1$ and an intrinsic line width of 30~eV at 6~keV as appropriate
for the low spin of T5X2. We scale the entire model for the factor of
2 difference in distance between J1808 and T5X2, and replace the J1808
column of $0.11\times 10^{22}$~cm$^{-2}$ with $1.5\times
10^{22}$~cm$^{-2}$ for T5X2 (Miller et al. 2012).  We simulate this
model, including the full NS surface features, through the SXS
response for 100ks. The count rate is 12 c/s so it is not piled up. 

We fit the simulated SXS data (Fig\ref{fig1}b) with a disk, blackbody
and comptonisation model (red, green and blue lines, respectively),
and there are clear residuals as shown in Fig \ref{fig1}c.  As in the
low inclination system Ser X-1 above, the largest features seen
correspond to the same 4 lines, SXVI at 4.7328\AA, a blend of
ArXVII/SXVI at mean wavelength of 3.9700\AA, CaXIX at 3.1773\AA and
FeXXV at 1.8508\AA. We fit four negative gaussians at these energies,
with redshift and broadening tied between them. While the SXVI line is
the most significant (Fig \ref{fig1}c), it has EW of only 5~eV.  The
FeXXV line by contrast has an EW of 10~eV but the detector is less
senstive at energies of $\sim 5$~keV than at $\sim 2$~keV, The fit
statistic decreases by $\chi^2=350$ for the addition of the four lines
so the derived redshift is well constrained at $z_\mathrm{surface}=0.3\pm
0.003$ and is unambiguous as it uses all 4 lines rather than relying
on the identification of a single feature.  The presence of other,
weaker features means that the broadening is slighly overestimated, at
$65\pm 10 $~eV at 6~keV.

There is a drop in the effective area of the SXS feature in the
response matrix at 2.18-2.3~keV. Sharp features in the response can
cause narrow residuals if the energy scale is not properly
calibrated. However, the edge in the current detector response is not
particularly sharp, so is unlikely to cause confusion with a narrow
line.

\begin{figure}[!b] 
\begin{center} 
\begin{tabular}{lll}
\includegraphics[scale=0.25]{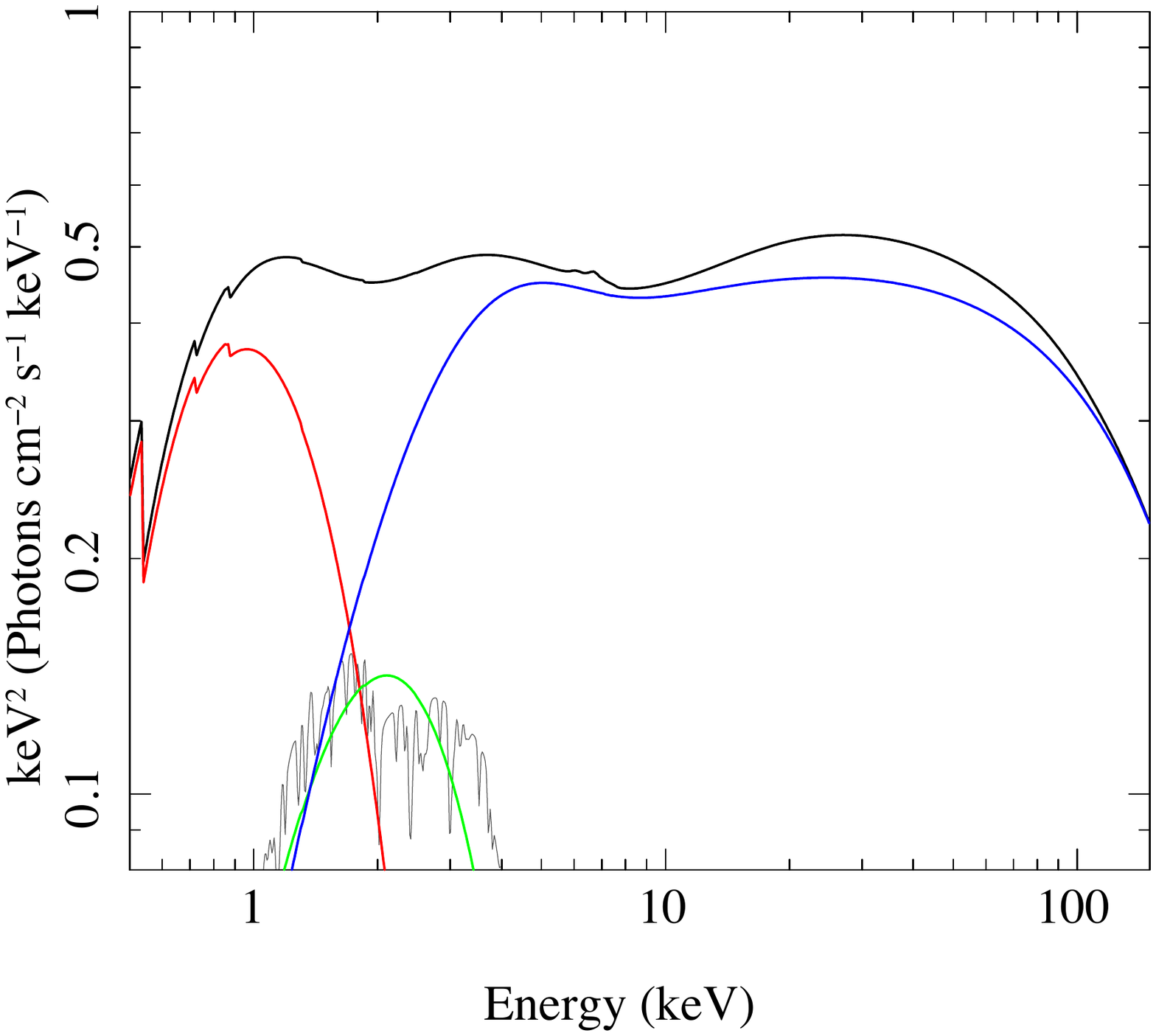} &
\includegraphics[scale=0.25]{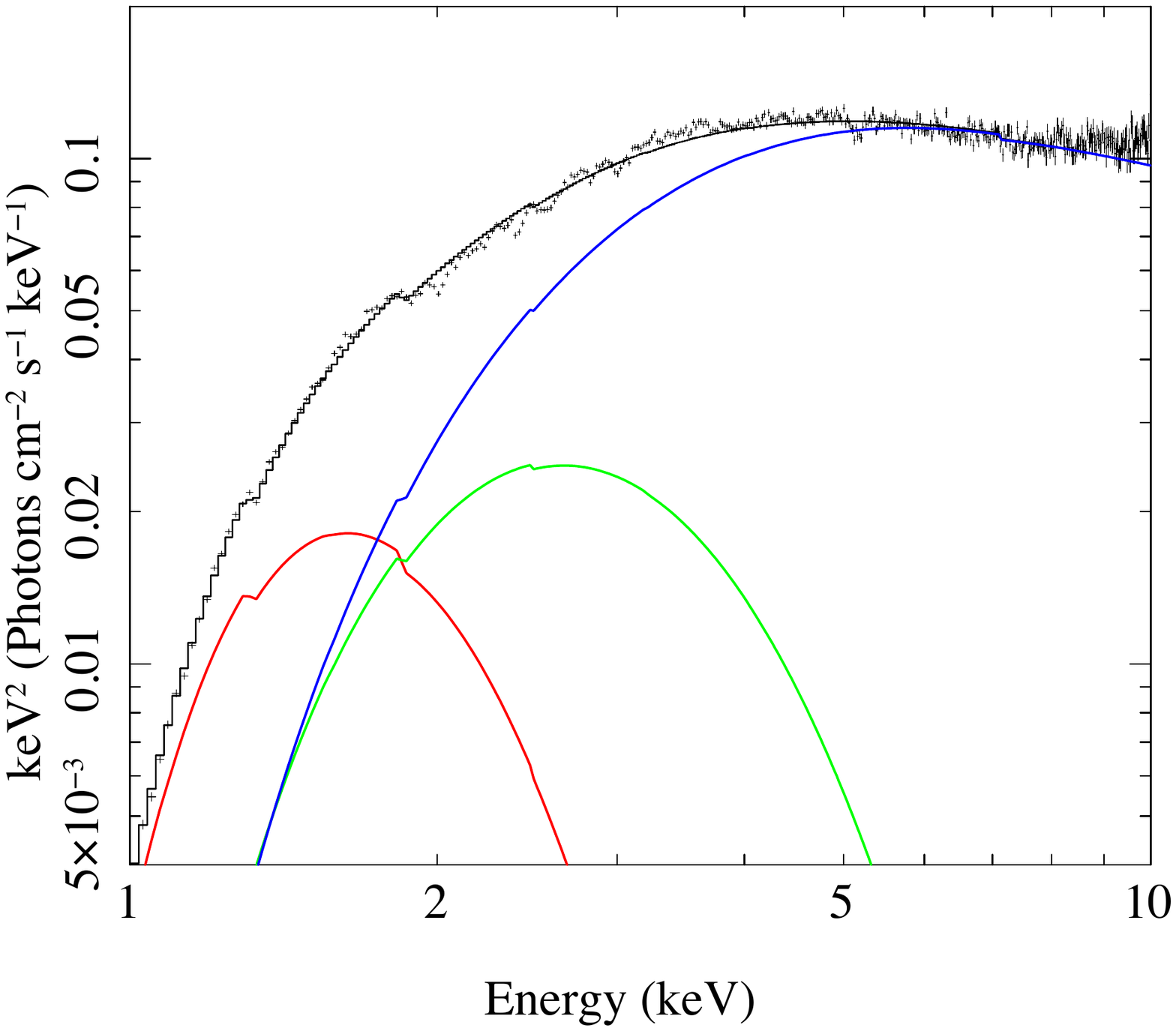} &
\includegraphics[scale=0.25]{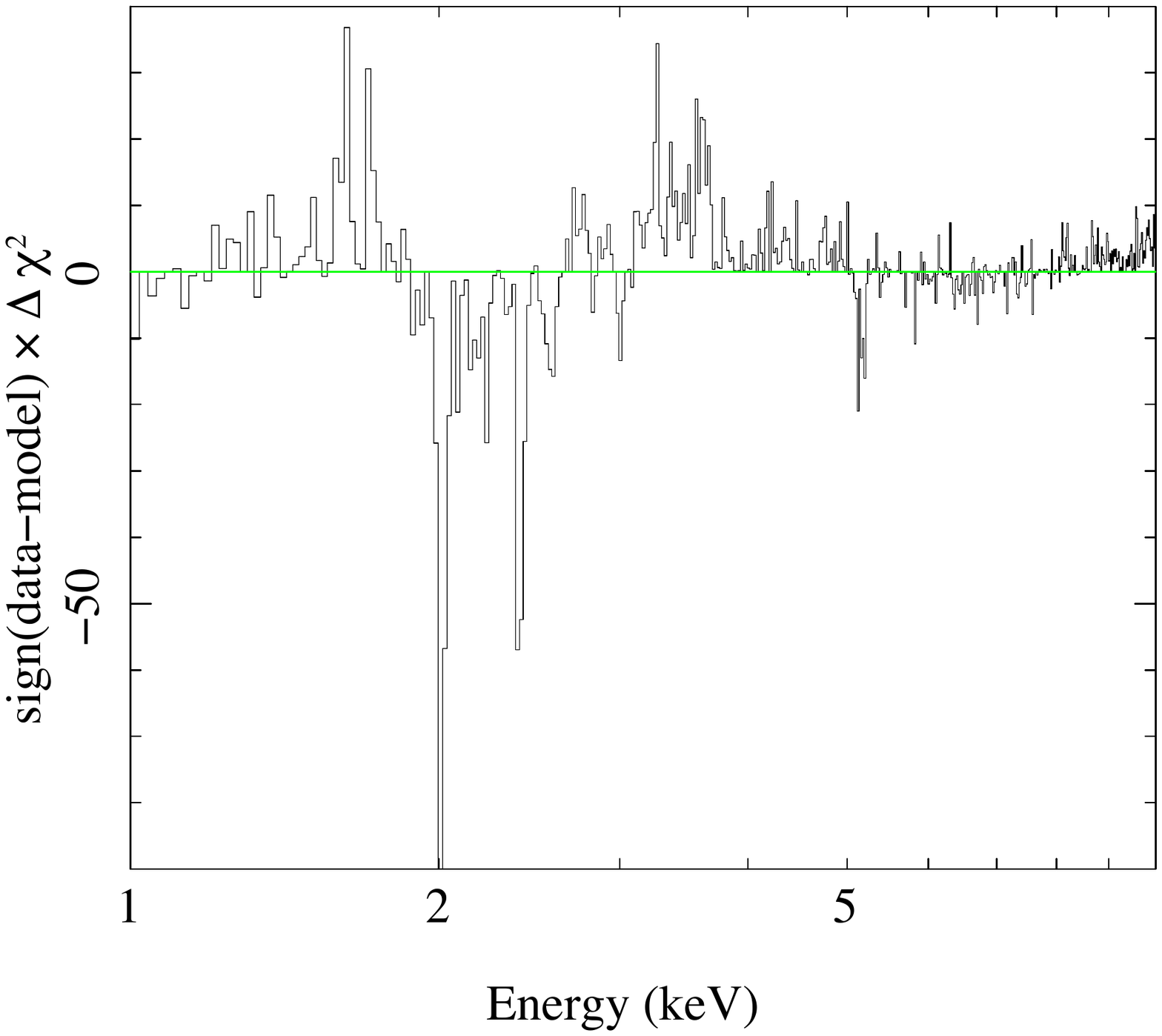}
\end{tabular}
\end{center}
\vspace{-32pt}
\caption{a) The model spectrum (black line) of an island state
observation of J1808 from Kajava et al. (2011). The red, green and blue
lines show the accretion disk, NS surface and corona, respectively.
The grey line shows an NS surface photosphere model, scaled to best
match the inferred surface emission in J1808. b) The model from a)
including the NS photosphere, scaled to the distance of T5X2 and with
$N_H=1.5\times 10^{22}$~cm$^{-2}$ simulated through the SXS response
for 100ks. This is fit with a continuum model with a disc (red) and
corona (blue) but the NS surface is asumed to be a blackbody (green),
i.e. without spectral features.  c) Residuals to the fit in b),
showing the absorption lines from SXVI ($\sim 2$~keV), ArXVII/SXVI
blend, CaXIX and FeXXV ($\sim 5$~keV).
Together these give an unambiguous, accurate measurement of $M/R$.}
\label{fig1} 
\end{figure}

\subsubsection{Beyond Feasibility: slow spin}

Long observations in the island state are almost
certain to sample X-ray bursts, with an average of 10 expected per
100ks observation (Galloway et al. 2008).  There is the possibility
that there are photospheric features in the burst spectra, especially
at the coolest temperatures seen which are $<1.5$~keV where iron
should not be completely ionised. Bursts from T5X2 give the
possibility of seeing these features (Blidstein, Chang \& Paerels
2003; Rauch, Suleimanov \& Werner 2008). The burst continuum
luminosity and temperature will also give constraints on M/R.

The T5X2 island state data show a low frequency (LF) QPO, which already
challenges the Lense-Thirring model for this feature as the very slow spin
in this system means that any torque between the spin and a misaligned
accretion flow is small (Altamirano et al. 2013). Nonetheless, the 
LF QPO signal can still be split up as a function of QPO phase to 
search for the `smoking gun' signature of precession of a vertically
tilted flow, where the iron line profile shifts bluer before the 
phase of peak intensity of the QPO, and
redder afterwards (Ingram \& Done 2012; see the {\it ASTRO-H} Black Holes white
paper: Miller et al. 2014).

Island states should also show reflection from the accretion disk. The
width of the broad line and the solid angle of the reflecting material
together can show whether or not the inner thin disk is replaced by a
hot flow as proposed in the truncated disk models. This can be
constrained by spectral fitting alone, but the continuum is complex,
with contributions from the disk, NS photosphere and Comptonisation
spectrum as well as the reflected emission from the disk and perhaps
some component of reflection from the NS surface (e.g. compare Cackett
et al. 2010 and Egron et al. 2012; see also Sanna et al. 2013). However,
the iron line and reflected continuum should also be lagged on a light
crossing time, so a combination of spectral-timing techniques (which
are currently being developed) may give better constraints.

\subsubsection{Phase resolved emission}

The accreting millisecond pulsars have pulsed emission showing that
the accretion flow is collimated onto the magnetic pole. Phase
resolved emission then limits the velocity broadening in each phase
bin. For a small shock this could recover a narrow feature which
shifted systematically with phase, but recent results indicate that
this shock region is probably fairly extended in azimuth (Kajava et al.
2011), so atomic features will still be broadened by the velocity
range from the NS surface even in a phase resolved anaysis, making
this unfeasible. All these systems are transients, so this would
additionally require a ToO

\section{Linking outflows to inflows in strong gravity: Winds in absorption}

\subsection{Background and Previous Studies}

The radius of the NS is similar to that of the last stable orbit, so
the gravitational potential seen by the accretion flow is the same as
that of black holes (BH). Thus the accretion flow structures (disk,
corona, disk jet, diskwinds) should be similar in both the NS and BH
systems, though the presence of a surface in the NS gives an
additional boundary layer continuum component, while an ergosphere
could (but probably doesn't: Migliari \& Fender 2006) give an
additional spin powered jet in the BH.

Absorption lines from ionised material are seen in many LMXB systems,
both NS and BH. These are most obvious during the deep absorption dips
connected to the accretion stream impact on the diskwhich modulate
high inclination sources on the orbital period, but there is also more
highly ionised material seen mostly via iron K$\alpha$ absorption
lines present outside of the dips, showing that it is distributed
above the disk at all azimuths (see e.g. the review by Diaz Trigo \&
Boirin 2012 for NS and Ponti et al. 2012 for BH). Key questions which
can be addressed with the SXS on {\it ASTRO-H} are a) How is this
material fed from the disk ?  b) How much mass is lost in the wind and
what is the impact of this on the accretion disk/jet structure ? c) Are
there any differences between the winds in BH and NS systems ?

Again, the SXS on {\it ASTRO-H} has enormous discovery
space in terms of the properties of this highly ionised material due
to its factor of 10 better energy resolution at iron K, as well as a
factor 10 increased effective area for sources below $\sim 4\times 10^{-9}$~ergs~cm$^{-2}$~s$^{-1}$
(pileup count rate limitations cut in above this flux).

\subsection{Prospects \& Strategy}

X-ray irradiation of the disk must produce a thermally driven
wind/corona. The X-ray flux from the innermost regions illuminates the
upper layers of the outer disk, heating it up to the Compton
temperature, $T_{\rm IC}$ (which depends only on spectral shape). The
heated upper layer expands on the sound speed $v_{IC}=(kT_{IC}/(\mu
m_p))^2$ due to the pressure gradient, so the thermal energy driving
the expansion is larger than the binding energy at radius $R_{IC}=
(1-L/L_{Edd})(v_{IC}/c)^{-2} R_g =6\times 10^5 (1-L/L_{Edd})
(kT_{ic}/10^7~K)^{-1} R_g=10^{12} (1-L/L_{Edd}) (kT_{ic}/10^7~K)^{-1}
(M/10M_\odot) $~cm so the material escapes as a wind (Begelman et al.\
1983). A more careful treatment shows that this leads to a wind from
the outer diskwhen $R>0.2R_{IC}$ and the luminosity is bright enough
to sustain rapid heating.  Conversely, at smaller radii/lower
luminosity the material remains bound but forms an extended atmosphere
above the disk (Begelman et al. 1983, Woods et al. 1996; Jimenez-Garate
et al. 2002). 

The strong X-ray illumination from the central source/inner accretion
disk ionises the wind, and the ionisation parameter $\xi=L/nR^2$ can
be estimated from the ratio of H to He-like absorption line equivalent
width (EW). However, these EW's are set by a combination of the column
density in that ion species, together with the turbulent velocity in
the wind. CCD resolution only gives an upper limit for the velocity
width of the line, which typically can be produced from a wide range
of column densities (saturated regime of the curve of growth).  This
uncertainty on column combines with uncertainty from the geometry
($N_H=n\Delta R < nR$) in estimating the launch radius $R<L/(\xi
N_H)$.

Current data from the NS systems show a good match to the thermal wind
predictions, as the absorbing material with inferred small radii
is static, while
outflows are only seen at radii larger than $0.2T_{IC}$ (Diaz Trigo \&
Boirin 2012).  This is in contrast to the controversy over the wind
driving mechanism in black hole LMXB systems. Many winds in the black
hole systems are also consistent with thermal driving, but the
dramatic absorption seen in one observation of GRO J1655-40 requires
additional mass loss, probably indicating magnetic driving (Miller et
al. 2006; Luketic et al. 2010) unless the wind is so thick as to cause
the source luminosity to be severely underestimated (Done, Gierlinski
\& Kubota 2007). The fact that such extreme winds are not seen from {\em
any} of the short period (small disk, persistent) high inclination
(dipping) NS systems (Diaz Trigo \& Boirin 2012) strongly argues
against any similar magnetic driving being important in the NS.  A
single observation of Cir X-1 is the only counterexample of an NS wind
outside of the thermal wind region (Diaz Trigo \& Boirin 2012), but
the spectral complexity often seen from this source means that its
luminosity could be strongly underestimated.

{\it ASTRO-H} will improve on this by giving a direct measure of the turbulent
velocity, to give a well constrained ion column from the observed
equivalent width. The improved sensitivity at high energy coupled with
the better energy resolution means that it is likely that the weak K$\beta$
absorption lines will be independently detected and constrained. 
These are not saturated so they give an unambiguous
measure of the ion column density. 

The mass loss due to a thermal wind increases with increasing disk
radius (Begelman et al. 1983; Woods et al. 1996), hence for the longest
period systems the prediction is that the mass loss rate is a factor
few larger than the observed mass accretion rate onto the compact
object. This mass loss rate can be measured from the ratio of emission
to absorption (P Cygni profile) seen in the wind lines. The best
current data from {\it Chandra} currenly only give an upper limit to the
solid angle subtended by the wind as less than a third of that
expected from a spherical wind in GX13+1 (Ueda et al. 2004). The much
better resolution of the SXS should enable a detection of the
associated emission, and hence give a direct measure of the mass loss
rate in the wind. 

In thermal wind models, the intrinsic line width is set mostly by the
velocity profile (decreasing with launch radius along the disk but
increasing with height as the material is accelerated by the thermal
expansion) rather than by true turbulence. The high resolution line
profiles of unsaturated lines should constrain this, allowing the
acceleration law to be reconstructed. The profiles should also
constrain the solid angle of the wind through the ratio of emission to
absorption, allowing the total mass loss rate to be determined. All
these features (launch radius, outflow velocity, velocity profile,
mass loss rate) can then be critically compared to the thermal wind
models to stringently test these against the first data which can
properly resolve the velocity structure.

\subsection{Targets \& Feasibility}

Many NS systems are physically small, so their disks do not extend far
enough to power a wind, so they can only show static coronae. However,
systems with evolved companions have larger orbits and hence larger
disks, so are more likely to show wind features. However, large disks
are also more likely to trigger the Hydrogen ionisation instability,
so these systems are transient unless the mass transfer rate from the
companion is very high (King \& Kolb 1999).  Data from the NS
systems show that the ionised material is significantly blueshifted
(i.e. is a wind) only where its inferred distance is larger than
$0.2R_{IC}$ (GX13+1, with very long orbital period of $\sim 24$~days,
and the transient T5X2), and that the outflow velocity is of order the
escape velocity at the launch radius, so $v_{IC}=(R/R_g)^{-1/2}c\sim
500-3000$~km~s$^{-1}$ for $0.2R_{IC}$, consistent with a thermal wind.

\begin{figure}
\begin{center} 
\begin{tabular}{lll}
\includegraphics[scale=0.25]{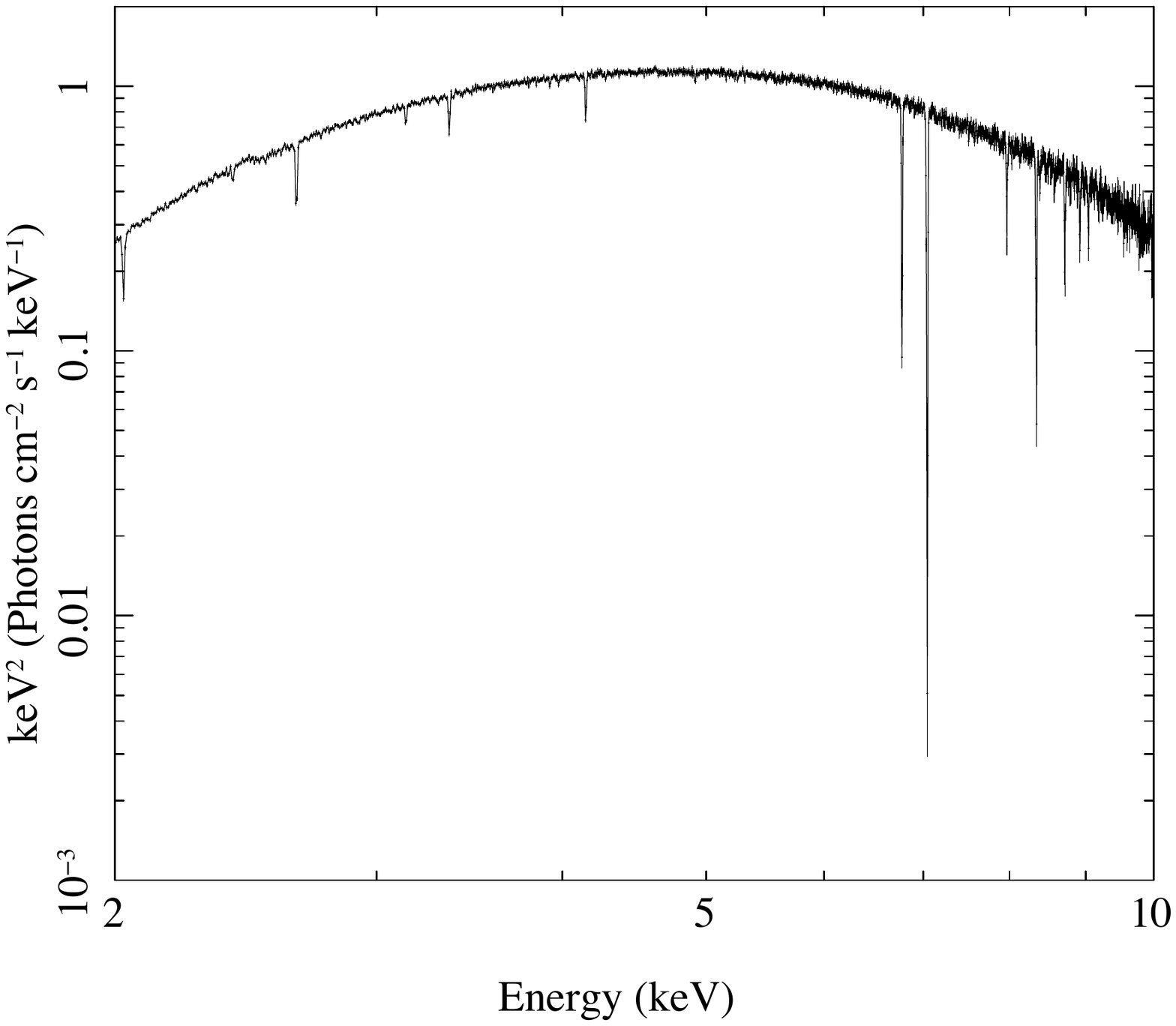} &
\includegraphics[scale=0.25]{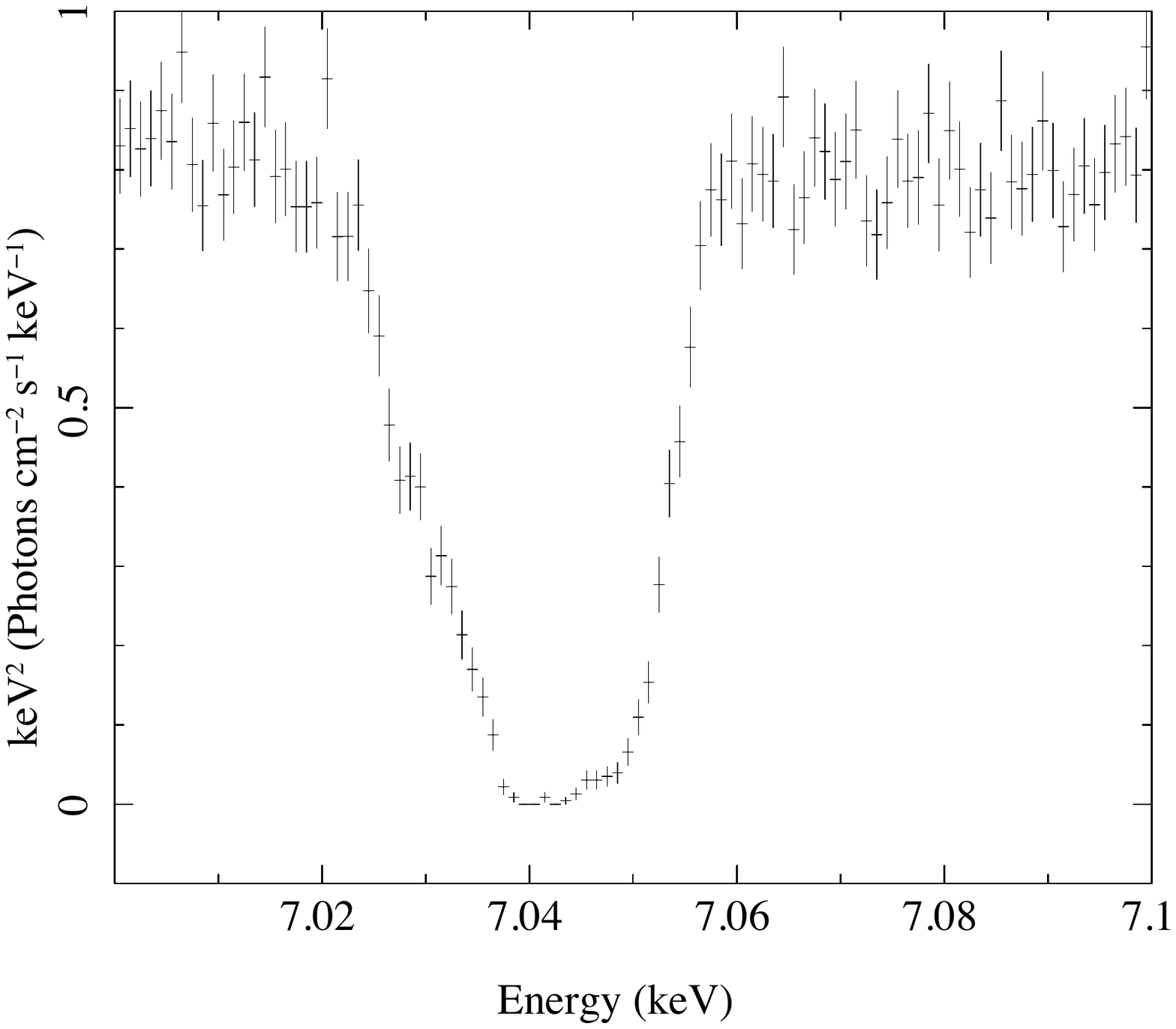} &
\includegraphics[scale=0.25]{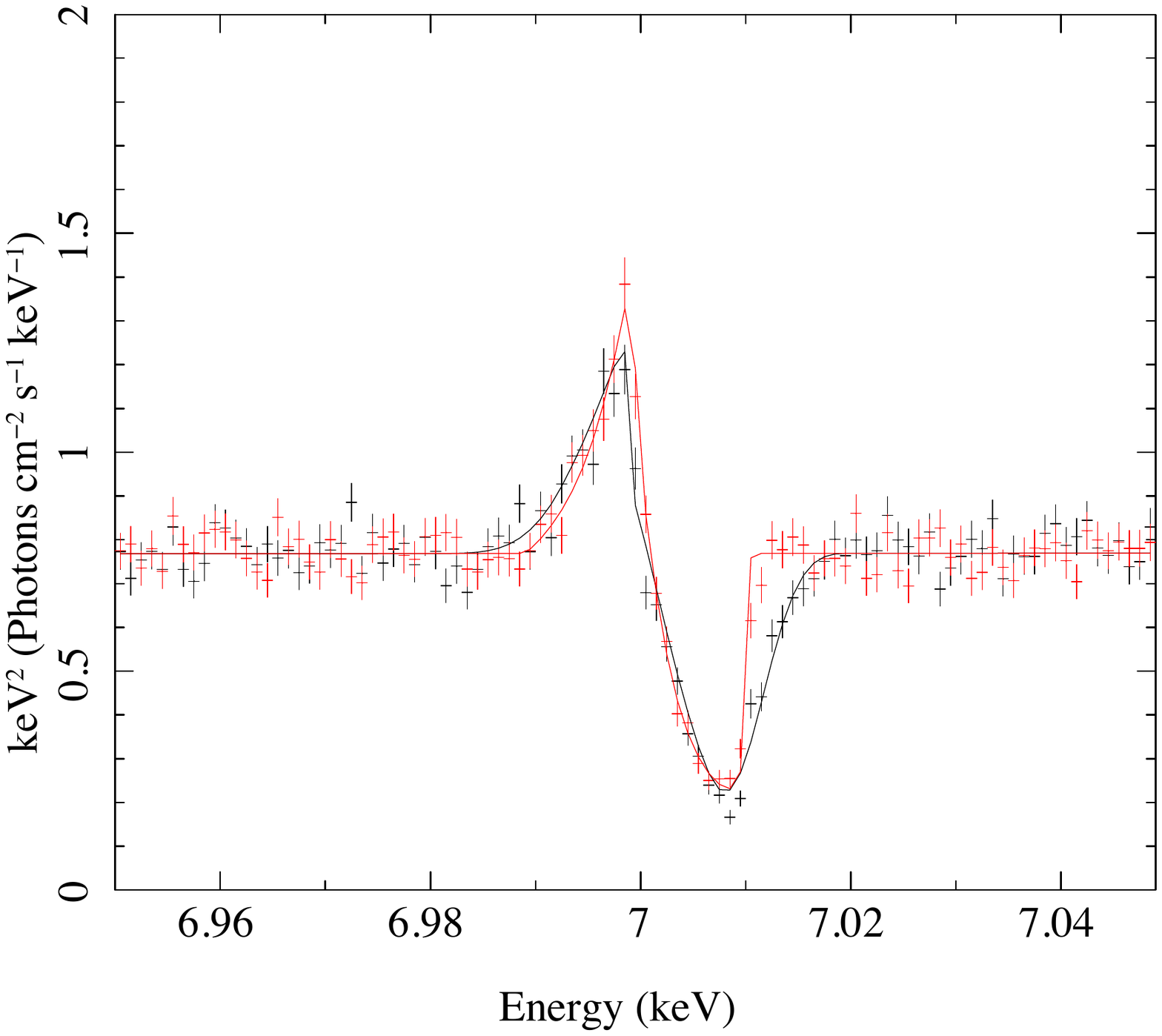}
\end{tabular} 
\end{center}
\vspace{-32pt}
\caption{a) The overall spectrum of GX13+1 simulated for 50~ks, similar
  to that seen with the {\it Chandra} HETG (Ueda et al. 2004). Multiple
  absorption lines are easily seen. b) Close up of the strongest Fe
  line from (a), showing that it is saturated. c) Red line:
  absorption/emission line P Cygni profile from a wind which
  accelerates quickly, so most material is at the terminal velocity,
  as expected for thermal winds. The black line is normalised to the
  same line shift and equivalent width, but has a much slower
  acceleration profile. The difference in the blue wing of the
  absorption is significantly detected.}
\label{fig2} 
\end{figure}

The LMXB catalog of Liu et al. (2007) shows that GX13+1 is the only
long period sub-Eddington (atoll rather than Z source) system.  It is
persistent, so scheduling is not an issue, and it persistently shows
absorption lines from Fe XXV and XXVI (Ueda et al. 2001; 2004; Diaz
Trigo et al. 2012), as well as lower equivalent width features from
Hydrogen-like Mn, Cr, Ca, Ar, S, Si, and Mg. The lines have a
blueshift of $\sim 450$~km s$^{-1}$ and an intrinsic width of $\sim
490\pm 110$~km s$^{-1}$ (i.e. at the limit of the {\it Chandra} HETG
resolution). The lines will be easily resolved with the
SXS. Fig \ref{fig2}a shows a simulation of a 50~ks SXS spectrum,
with flux reduced by a factor 3 to give a countrate of 65 c/s. 
The model {\tt tbabs$\times$zxipcf$\times$(diskbb+bbody+gau)} has an ionised
absorber with column of $3\times 10^{23}$ cm$^{-2}$ and $\log
\xi=4.41$ (similar to that seen by Ueda et al. 2004). 
Fig \ref{fig2}b shows a close up of the resolved line profile around Fe XXVI
K$\alpha$. This ionised absorber was simulated with a 
turbulent velocity of 200~km/s and is easily resolved by the
SXS, and is clearly saturated, being completely black in the
center. None of the other lines are saturated, so the column can be
measured unambiguously from the Fe XXVI K$\beta$ line. 

However, winds are not dominated by turbulent velocities. Instead,
their velocity structure is determined by the acceleration mechanism,
imprinting itself onto the (unsaturated) line profiles. Standard
P-Cygni lines can be fit assuming a simple model for the column at
each velocity of $\tau(v)\propto (1-v/v_\infty)^\alpha$.  The red and
black lines in Fig \ref{fig2}c show what happens to the line shapes for
different acceleration laws, parameterised by different $\alpha$ and
$v_\infty$. Thermals winds have the majority of the acceleration
happen close to the launch radius, so most of the material is
outflowing at the terminal velocity 
(red line: $\alpha=0.2$ with $v_\infty=450$~km/s). By
contrast, magnetic driving could give rise to much slower
acceleration, where the majority of the absorption occurs for material
which is not yet at its terminal velocity (black line: $\alpha=4$ with
$v_\infty=900$~km/s scaled to the same depth). The change in shape of
the blue wing of the absorption line can be seen in a 100~ks {\it ASTRO-H}
simulation (red and black points) of a source with 2-10~keV count rate
of 45 counts/s in the SXS, though it may be possible to
recover this significance from a 50~ks observation by 
stacking the line profiles from all
unsaturated lines. These different velocity
structures give the most sensitive diagnostic of the launch mechanism.

The models in Fig \ref{fig2}c were for a pure P-Cygni profile (Lamers et al.
1987; Done et al. 2007) i.e. assuming a spherically symmetric winds, so
give maximum emission associated with the absorption. The {\it Chandra}
HETG already limits the solid angle of the wind to be less than a
third of this maximum (Ueda et al. 2004), so the emission in these
profiles is overestimated. Fitting the detailed line profiles should 
determine the ratio of emission to absorption, and hence give an
unambiguous measure of the solid angle of the wind as required to
determine the total mass loss rate. More theoretical work to develop
P-Cygni profiles from diskwinds should be done so that these models
can explicitally be fit to the data to constrain the solid angle of
the wind. 

\subsection{Beyond Feasibility}

GX13+1 is in the upper banana branch, so should have its disk be close
to the NS surface. The broad iron line is clearly seen
in the {\it XMM-Newton} datasets (Diaz-Trigo et al. 2012). The width of
the line should determine the inner radius of the disk, and hence give
an upper limit to the size of the NS surface. 

This source also shows weak evidence for a hard tail above 50~keV from
{\it INTEGRAL} data (Paizis et al. 2006). The
HXT and SGD will be able to constrain this component with much greater
confidence, so that its nature and origin can be better understood.

\section{Linking outflows to inflows in strong gravity: Winds in emission}

\subsection{Background and Previous Studies}

LMXB seen at very high inclination have the inner disk completely
obscured by the outer disk so the source is seen only indirectly
through scattering in the corona and wind, though this reflected
emission is also overlaid by recombination from the photoionised
corona and wind. These are called accretion disk corona (ADC) sources,
and they give a different angle on the corona and wind structure above
the disk than seen in the absorption lines discussed above.

\subsection{Prospects \& Strategy}

\begin{figure}
\begin{center} 
\includegraphics[width=0.9\hsize]{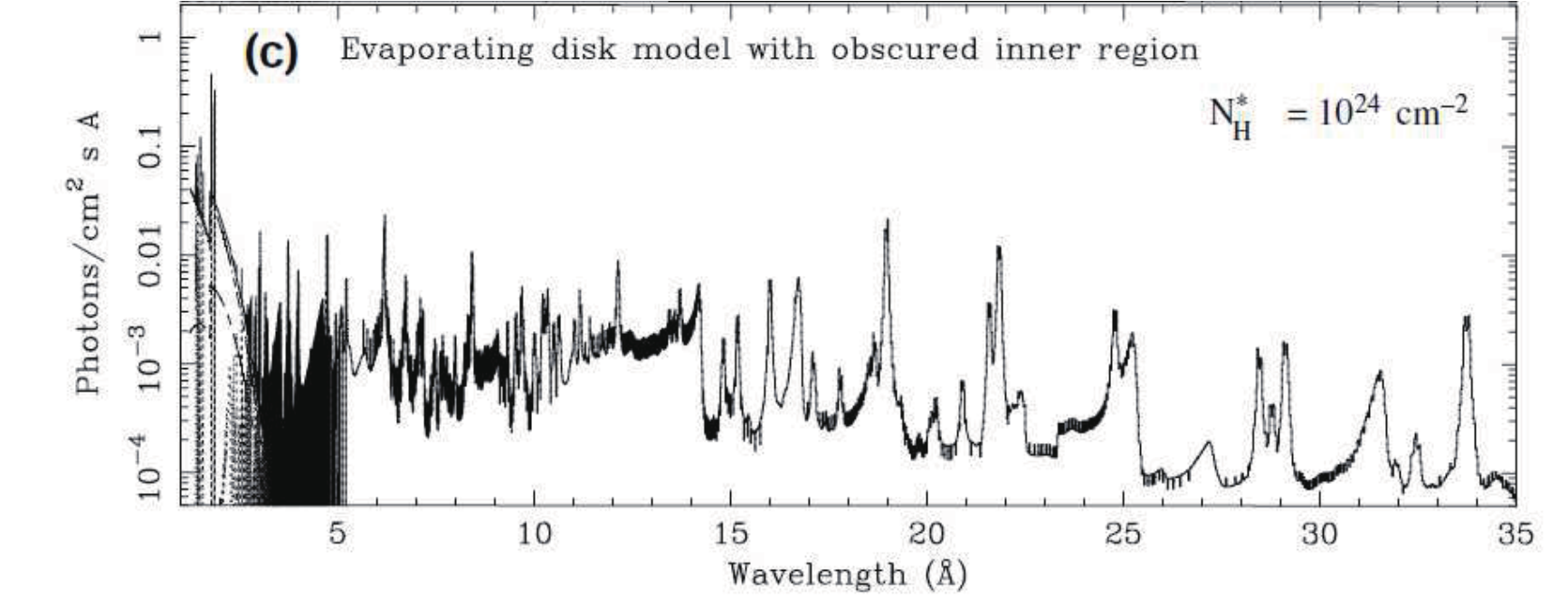}
\end{center}
\vspace{-12pt}
\caption{The recombination line and continuum spectrum from a corona
  above a NS disk where the central NS and disk is obscured by a
  column of $10^{24}$ cm$^{-2}$ (Jimenez-Garate et al. 2002)
}
\label{fig3} 

\end{figure}

The spectrum of emission lines from the X-ray heated wind and/or
corona above the disk is extremely rich in plasma diagnostics
(e.g. the theoretical calculations of the corona by Jimenez-Garate et
al. 2002 shown in Fig \ref{fig3}). 

\subsection{Targets \& Feasibility}

To compare the wind features (not just the static corona) requires
that we use an ADC which has a large enough orbital period to drive a
wind. This uniquely selects 2S0921-63, the longest period ADC source
(9.02 days e.g. Kallman et al. 2003). This shows true eclipses as well
as orbital ADC modulation, and the eclipse gives the cleanest possible
diagnostic of the scattering material at large distances from the
source, weighting the emission to the wind region rather than the
static corona below $0.2R_{IC}$. The eclipse lasts up to 80~ks, setting
the upper limit to the time for a single observation, with 2-10~keV
flux of $9\times 10^{-11}$ ergs~cm$^{-2}$~s$^{-1}$ (EPIC PN) with
strong line emission from Fe XXV and XXVI as well as multiple species
at lower energy (Kallman et al. 2003). The iron K$\alpha$ (and $\beta$)
line intrinsic widths should uniquely determine the wind outflow
velocity. These emission lines should also be accompanied by by
absorption lines, though these are much reduced due to the geometry.
These intrinsic line widths are not well constrained by the EPIC PN
and MOS cameras due to their poor resolution, with width $<1300$ km/s,
nor are they well constrained by the {\it Chandra} HETG data due to low
statistics.

\begin{figure}
\begin{center} 
\includegraphics[width=0.6\hsize]{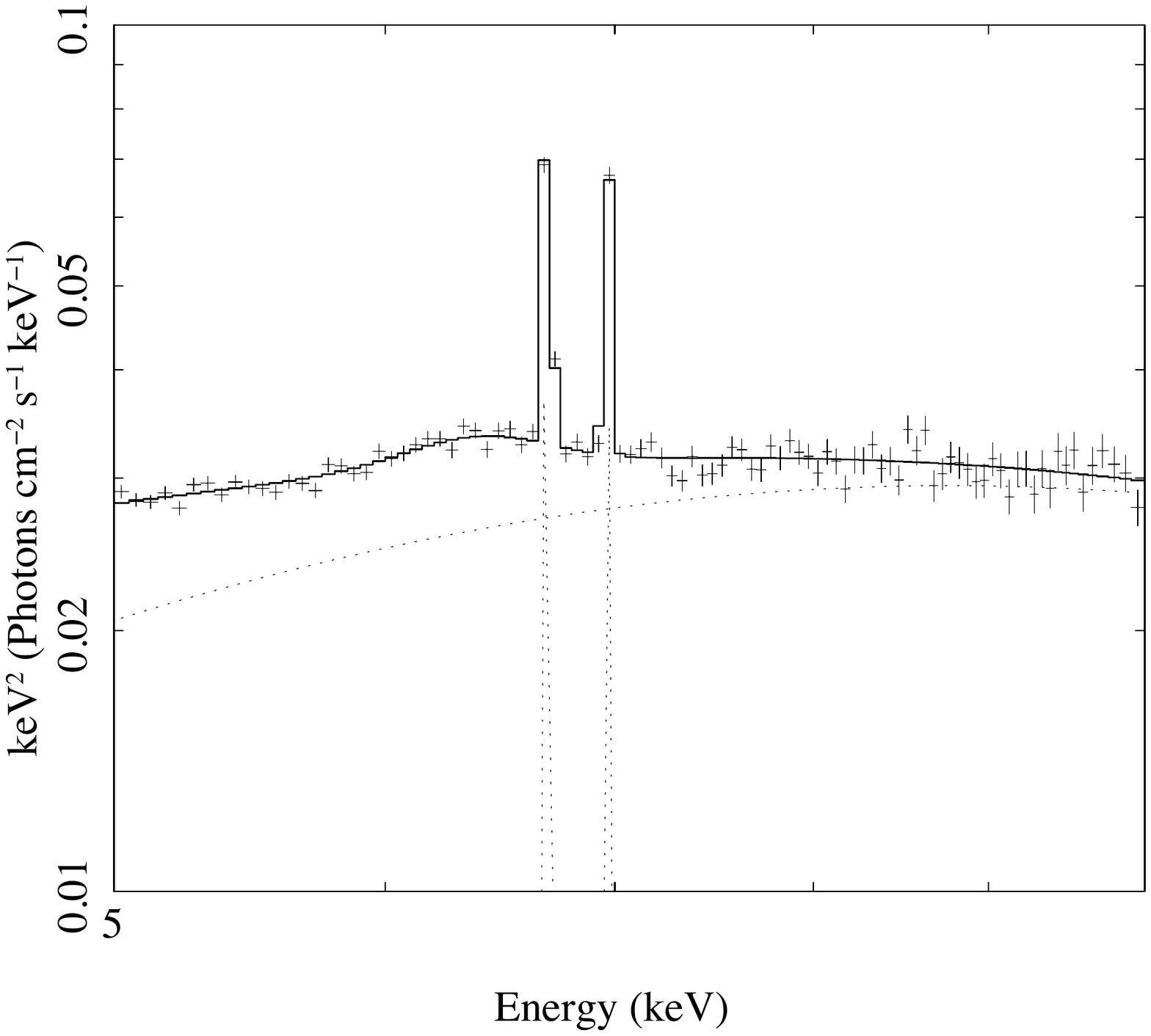}
\end{center}
\vspace{-12pt}
\caption{Simulated 80~ks eclipse spectrum of 2S0921-63. The data can
  easily determine the intrinsic width of the Fe XXV and XXVI emission
  lines, and hence constrain the velocity of the scattering material.
}
\label{fig4} 

\end{figure}

We simulate an 80~ks eclipse spectrum using the {\it XMM-Newton} data reported in
Kallman et al. (2003) giving 4 cts/s in the SXS (Fig
\ref{fig4}). The Fe K$\alpha$ lines are simulated assuming these are
from a wind with outflow velocity of 450 km/s, giving an intrinsic
width of 450km/s (i.e. 0.01~keV). This intrinsic width of both 6.7 and
6.95~keV emission lines can be easily constrained in the 80~ks
simulation, showing that this is detecting emission from the wind.

The alternative bright ADC source X1822-371 has much shorter orbital
period of 5.57 hours. This gives an implied disk size of $\sim 5\times
10^{10}$~cm, which is just large enough for a thermal wind to form,
especially as this system is probably intrinsically close to
$L_{Edd}$. There are existing {\it Chandra} HETG spectra of this source
which show a weak FeXXVI resonance line which is consistent with
comming from the static corona/outflow region with $R<3.7\times
10^{10}$~cm (from eclipse location). This emission line is narrow,
with $\sigma<400$ km/s (Ji et al. 2011; Iaria et al. 2013), implying
that the wind outflow velocity is of this order. This 
seems surprisingly small for a system close to $L_{Edd}$.
Most of the other lower ionisation lines come from material associated
with the disk edge (Ji et al. 2011; Iaria et al. 2013).

\subsection{Beyond Feasibility}

Bursts are unlikely to be seen as the source is intrinsically quite
luminous, so should be in the regime of stable nuclear burning. 
However, the lower energy line emission spectrum should be extremely rich,
allowing multiple other diagnostics of the wind and corona. 

\section{Spectral modelling requirements}

There is still much work to be done on model development in order to
access the full science from SXS observations. These are:

Section 1 on the NS equation of state. The models of the NS surface
spectra have so far only been calculated for purely thermal
emission. However, the island state has a strong hard X-ray component
which will illuminate the surface, potentially changing the gradient
in ionisation state with scale height. These effects should be
incorporated into the models in order to get a better estimate of the
predicted equivalent width of the absorption lines. However, any
significant detection of these absorption lines is a clean measure of
$M/R$ so this is not required to interpret the data if the lines are
seen.  More fundamentally, models of the abundance of the NS surface
during steady accretion (rather than bursts) are highly uncertain, and
it is possible that the heavy elements do sink below the photosphere
so that there are no atomic features from the surface at all.

Section 2 on the wind in absorption from LMXRB. 
The wind absorption will likely be first characterised using the XSTAR
photoionisation absorption models. These have recently been extended
to include the doublet structure of the K$\alpha$ line from H-like
iron, however the data will also require this for He-like iron as
well. 

The emission associated with the wind cannot be calculated just using
the (spherical) P Cygni approach as the wind is equatorial. Better
models which include the emission and absorption for a given geometry
are needed to get the tightest constraints from the observed line
profiles. Updated (Woods et al. 1996) hydrodynamic simulations of
thermal winds including radiation transfer are needed to make 
detailed predictions of the absorption line profiles predicted by
thermal wind models. 

Section 3 on the wind in emission from ADC LMXRB. There are
sophisticated models, but there needs to be a lot of development to
bridge between the models and diagnostics of the wind versus static
corona and the dipping material. 

\section{Summary of Top Science}

\begin{itemize}
\item[1.] {\bf Objective:} Measure $M/R$ of a NS to a few
  percent.
\item[] {\bf Why:} To constrain the equation of state of dense matter
  to order of magnitude better accuracy than the 10-20\% from current
  measurements.
\item[] {\bf How:} By detecting absorption lines from the surface of a
neutron star with low projected velocity from rotation. This requires
either low inclination (the steady source Ser X-1, on the banana
branch where the NS surface at the poles should be seen directly) 
or slow rotation (the transient source T5X2, an 11~Hz AMXP, where the
surface should be directly visible in the island state, requiring a ToO). 
\bigskip

\item[2.] {\bf Objective:} To measure the velocity profile of the
  accretion disk wind
\item[] {\bf Why:} To determine how the wind is launched and accelerated.
\item[] {\bf How:} Thermal wind models make some very specific
  predictions about the radial structure of the wind - its launch
  radius and its velocity profile. These are testable from high
  signal-to-noise SXS observations of the H and He-like
  absorption (and emission) lines in both K$\alpha$ and $\beta$. 
\end{itemize}


\clearpage
\begin{multicols}{2}
\end{multicols}


{\footnotesize

\begin{thebibliography}{99}

\bibitem[Altamirano et al.(2010)]{2010ATel.2932....1A} Altamirano, D., 
Watts, A., Kalamkar, M., et al.\ 2010, The Astronomer's Telegram,
2932, 1 
\bibitem[Altamirano et al.(2010)]{2010ATel.2952....1A} Altamirano, D., 
Homan, J., Linares, M., et al.\ 2010, The Astronomer's Telegram, 2952, 1 
\bibitem[Altamirano et al.(2012)]{2012ApJ...759L..20A} Altamirano, D., 
Ingram, A., van der Klis, M., et al.\ 2012, Ap.J. Letts., 759, L20 
\bibitem[Barret(2012)]{2012ApJ...753...84B} Barret, D.\ 2012, Ap.J., 753, 84 
\bibitem[Baub{\"o}ck et al.(2013)]{2013ApJ...766...87B} Baub{\"o}ck, M., 
Psaltis, D., Ozel, F.\ 2013, ApJ, 766, 87 
\bibitem[Begelman et al.(1983)]{1983ApJ...271...70B} Begelman, M.~C., 
McKee, C.~F., \& Shields, G.~A.\ 1983, Ap.J., 271, 70 
\bibitem[Bildsten et al.(2003)]{2003ApJ...591L..29B} Bildsten, L., Chang, 
P., \& Paerels, F.\ 2003, Ap.J.Letts, 591, L29 
\bibitem[Cackett et al.(2010)]{2010ApJ...720..205C} Cackett, E.~M., Miller, 
J.~M., Ballantyne, D.~R., et al.\ 2010, ApJ, 720, 205 
\bibitem[Cavecchi et al.(2011)]{2011ApJ...740L...8C} Cavecchi, Y., Patruno, 
A., Haskell, B., et al.\ 2011, ApJL, 740, L8 
\bibitem[Chakraborty et al.(2011)]{2011MNRAS.418..490C} Chakraborty, M., 
Bhattacharyya, S., \& Mukherjee, A.\ 2011, MNRAS, 418, 490 
\bibitem[Chang et al.(2005)]{2005ApJ...629..998C} Chang, P., Bildsten, L., 
\& Wasserman, I.\ 2005, Ap.J., 629, 998 
\bibitem[Cornelisse et al.(2013)]{2013MNRAS.432.1361C} Cornelisse, R., 
Casares, J., Charles, P.~A., \& Steeghs, D.\ 2013, MNRAS, 432, 1361 
\bibitem[Cumming et al.(2001)]{2001ApJ...557..958C} Cumming, A., Zweibel, 
E., \& Bildsten, L.\ 2001, ApJ, 557, 958 
\bibitem[Diaz Trigo 
\& Boirin(2012)]{2012arXiv1210.0318D} Diaz Trigo, M., \& Boirin, L.\ 2012, arXiv:1210.0318 
\bibitem[D{\'{\i}}az Trigo et 
al.(2012)]{2012A&A...543A..50D} D{\'{\i}}az Trigo, M., Sidoli, L.,
  Boirin, L., \& Parmar, A.~N.\ 2012, A\& A, 543, A50 
\bibitem[Done et 
al.(2007)]{2007A&ARv..15....1D} Done, C., Gierli{\'n}ski, M., \&
  Kubota, A.\ 2007, ARA\& A, 15, 1 
\bibitem[Done et al.(2007)]{2007MNRAS.374L..15D} Done, C., Sobolewska, 
M.~A., Gierli{\'n}ski, M., \& Schurch, N.~J.\ 2007, MNRAS, 374, L15 
\bibitem[Egron et 
al.(2013)]{2013A&A...550A...5E} Egron, E., Di Salvo, T., Motta, S., et
  al.\ 2013, A\& A, 550, A5 
\bibitem[Frank et al.(2002)]{2002apa..book.....F} Frank, J., King, A., 
\& Raine, D.~J.\ 2002, Accretion Power in Astrophysics, by Juhan Frank and Andrew King and Derek Raine, pp.~398.~ISBN 0521620538.~Cambridge, UK: Cambridge University Press, February 2002.
\bibitem[Galloway et al.(2008)]{2008ApJS..179..360G} Galloway, D.~K., Muno, 
M.~P., Hartman, J.~M., Psaltis, D., 
\& Chakrabarty, D.\ 2008, Ap.J.Supp., 179, 360 
\bibitem[Gierli{\'n}ski et al.(2002)]{2002MNRAS.331..141G} Gierli{\'n}ski, 
M., Done, C., \& Barret, D.\ 2002, MNRAS, 331, 141 
\bibitem[Hebeler et al.(2013)]{2013ApJ...773...11H} Hebeler, K., Lattimer, 
J.~M., Pethick, C.~J., \& Schwenk, A.\ 2013, ApJ, 773, 11 
\bibitem[Iaria et 
al.(2013)]{2013A&A...549A..33I} Iaria, R., Di Salvo, T., D'A{\`i}, A.,
  et al.\ 2013, A\& A, 549, A33 
\bibitem[Ingram 
\& Done(2012)]{2012MNRAS.427..934I} Ingram, A., \& Done, C.\ 2012, MNRAS, 427, 934 
\bibitem[Ji et al.(2011)]{2011ApJ...729..102J} Ji, L., Schulz, N.~S., 
Nowak, M.~A., \& Canizares, C.~R.\ 2011, Ap.J., 729, 102 
\bibitem[Jimenez-Garate et al.(2002)]{2002ApJ...581.1297J} Jimenez-Garate, 
M.~A., Raymond, J.~C., \& Liedahl, D.~A.\ 2002, Ap.J., 581, 1297 
\bibitem[Kallman et al.(2003)]{2003ApJ...583..861K} Kallman, T.~R., 
Angelini, L., Boroson, B., \& Cottam, J.\ 2003, Ap.J., 583, 861 
\bibitem[Kajava et al.(2011)]{2011MNRAS.417.1454K} Kajava, J.~J.~E., 
Ibragimov, A., Annala, M., Patruno, A., 
\& Poutanen, J.\ 2011, MNRAS, 417, 1454 
\bibitem[King et al.(1996)]{1996ApJ...464L.127K} King, A.~R., Kolb, U., 
\& Burderi, L.\ 1996, ApJL, 464, L127 
\bibitem[King \& Kolb(1999)]{1999MNRAS.305..654K} King, A.~R., \&
Kolb, U.\ 1999, MNRAS, 305, 654
\bibitem[Lamb et al.(2009)]{2009ApJ...706..417L} Lamb, F.~K., Boutloukos, 
S., Van Wassenhove, S., et al.\ 2009, ApJ, 706, 417 
\bibitem[Lamers et al.(1987)]{1987ApJ...314..726L} Lamers, H.~J.~G.~L.~M., 
Cerruti-Sola, M., \& Perinotto, M.\ 1987, ApJ, 314, 726 
\bibitem[Lin et al.(2009)]{2009ApJ...696.1257L} Lin, D., Remillard, R.~A., 
\& Homan, J.\ 2009, ApJ., 696, 1257 
\bibitem[Loeb(2003)]{2003PhRvL..91g1103L} Loeb, A.\ 2003, Physical Review 
Letters, 91, 071103 
\bibitem[Liu et 
al.(2007)]{2007A&A...469..807L} Liu, Q.~Z., van Paradijs, J., \& van
  den Heuvel, E.~P.~J.\ 2007, A\& Ap, 469, 807 
\bibitem[Luketic et al.(2010)]{2010ApJ...719..515L} Luketic, S., Proga, D., 
Kallman, T.~R., Raymond, J.~C., \& Miller, J.~M.\ 2010, Ap.J., 719,
515 
\bibitem[Medvedev 
\& Narayan(2001)]{2001ApJ...554.1255M} Medvedev, M.~V., \& Narayan,
  R.\ 2001, ApJ, 554, 1255 
\bibitem[Migliari 
\& Fender(2006)]{2006MNRAS.366...79M} Migliari, S., \& Fender, R.~P.\ 2006, MNRAS, 366, 79 
\bibitem[Miller et al.(2006)]{2006Natur.441..953M} Miller, J.~M., Raymond, 
J., Fabian, A., et al.\ 2006, Nature, 441, 953 
\bibitem[Miller et al.(2011)]{2011ApJ...731L...7M} Miller, J.~M., Maitra, 
D., Cackett, E.~M., Bhattacharyya, S., 
\& Strohmayer, T.~E.\ 2011, ApJL, 731, L7 
\bibitem[Miller et al.(2013)]{2013ApJ...779L...2M} Miller, J.~M., Parker, 
M.~L., Fuerst, F., et al.\ 2013, ApJL, 779, L2 
\bibitem[{\"O}zel(2013)]{2013RPPh...76a6901O} {\"O}zel, F.\ 2013, Reports 
on Progress in Physics, 76, 016901 
\bibitem[Papitto et al.(2012)]{2012MNRAS.423.1178P} Papitto, A., Di Salvo, 
T., Burderi, L., et al.\ 2012, MNRAS, 423, 1178 
\bibitem[Papitto et al.(2013)]{2013Natur.501..517P} Papitto, A., Ferrigno, 
C., Bozzo, E., et al.\ 2013, Nature, 501, 517 
\bibitem[Patruno 
\& Watts(2012)]{2012arXiv1206.2727P} Patruno, A., \& Watts, A.~L.\
  2012, arXiv:1206.2727 
\bibitem[Paizis et 
al.(2006)]{2006A&A...459..187P} Paizis, A., Farinelli, R., Titarchuk,
  L., et al.\ 2006, A\& A, 459, 187 
\bibitem[Ponti et al.(2012)]{2012MNRAS.422L..11P} Ponti, G., Fender, R.~P., 
Begelman, M.~C., et al.\ 2012, MNRAS, 422, L11 
\bibitem[Rauch et 
al.(2008)]{2008A&A...490.1127R} Rauch, T., Suleimanov, V., \& Werner,
  K.\ 2008, A\& A, 490, 1127 
\bibitem[Revnivtsev et al.(2013)]{2013MNRAS.434.2355R} Revnivtsev, M.~G., 
Suleimanov, V.~F., \& Poutanen, J.\ 2013, MNRAS, 434, 2355 
\bibitem[Sakurai et al.(2014)]{2014PASJ...66...10S} Sakurai, S., Torii, S., 
Noda, H., et al.\ 2014, PASJ, 66, 10 
\bibitem[Sanna et al.(2013)]{2013MNRAS.432.1144S} Sanna, A., Hiemstra, B., 
M{\'e}ndez, M., et al.\ 2013, MNRAS, 432, 1144 
\bibitem[Suleimanov 
\& Poutanen(2006)]{2006MNRAS.369.2036S} Suleimanov, V., \& Poutanen, J.\ 2006, MNRAS, 369, 2036 
\bibitem[Suleimanov et 
al.(2012)]{2012A&A...545A.120S} Suleimanov, V., Poutanen, J., \&
  Werner, K.\ 2012, A\& Ap, 545, A120 
\bibitem[Ueda et al.(2001)]{2001ApJ...556L..87U} Ueda, Y., Asai, K., 
Yamaoka, K., Dotani, T., \& Inoue, H.\ 2001, Ap.J.Letts, 556, L87 
\bibitem[Ueda et al.(2004)]{2004ApJ...609..325U} Ueda, Y., Murakami, H., 
Yamaoka, K., Dotani, T., \& Ebisawa, K.\ 2004, Ap.J., 609, 325 
\bibitem[Woods et al.(1996)]{1996ApJ...461..767W} Woods, D.~T., Klein, 
R.~I., Castor, J.~I., McKee, C.~F., \& Bell, J.~B.\ 1996, Ap.J., 461, 767 


\end{thebibliography}

}
\end{document}